\documentclass[12pt,preprint]{aastex}
\def\lesssim{\mathrel{\hbox{\rlap{\hbox{\lower4pt\hbox{$\sim$}}}\hbox{$<$}}}}
\def\gtrsim{\mathrel{\hbox{\rlap{\hbox{\lower4pt\hbox{$\sim$}}}\hbox{$>$}}}}

\newcommand{\etal  }{{et al.} } 
\newcommand{\msun}{\thinspace M_\odot} 
\newcommand{\feoh}{[{\rm Fe} / {\rm H}]} 
\shorttitle{Low-Mass Star Formation Triggered by the First Supernovae}
\shortauthors{Machida \etal 2003}

\begin{document}

\title{Low-Mass Star Formation, Triggered by Supernova in Primordial Clouds}

\author{Masahiro N. Machida\altaffilmark{1} } 
\affil{Center for Frontier Science, 
Chiba University, Yayoicho 1-33, Inageku, Chiba 263-8522, Japan}

\author{Kohji Tomisaka\altaffilmark{2}} 
\affil{National Astronomical Observatory, Mitaka, Tokyo 181-8588, Japan}

\author{Fumitaka Nakamura\altaffilmark{3}}
\affil{Faculty of Education and Human Sciences, Niigata University, Ikarashi 2-8050,Niigata 950-2181,Japan}

\author{Masayuki Y. Fujimoto\altaffilmark{4}}
\affil{Department of Physics, Hokkaido University, Sapporo 060-0810, Japan}

\altaffiltext{1}{machida@cfs.chiba-u.ac.jp}
\altaffiltext{2}{tomisaka@th.nao.ac.jp}
\altaffiltext{3}{fnakamur@ed.niigata-u.ac.jp}
\altaffiltext{4}{fujimoto@astro1.sci.hokudai.ac.jp}

\begin{abstract}

The evolution of a gas shell, swept by the supernova remnant of a massive first generation star, is studied with H$_2$ and HD chemistry taken into account and with use of a semi-analytical approximation to the dynamics.  
When a first-generation star, formed in a parent pregalactic cloud, explodes as a supernova with explosion energy in the range of $10^{51}{\rm erg}- 10^{52}$ erg at redshifts of $z = 10 - 50$, H$_2$ and HD molecules are formed in the swept gas shell at fractional abundances of $\sim  10^{-3}$ and $\sim 10^{-5}$, respectively, and effectively cool the gas shell to temperatures of $32 {\rm K} - 154$ K.  
If the supernova remnant can sweep to gather the ambient gas of mass $6 \times 10^4 \msun - 8 \times 10^5 \ M_\odot$, the gas shell comes to be dominated by its self-gravity, and hence, is expected to fragment. 
The amount of swept gas necessary for fragmentation increases with the explosion energy and decreases with the interstellar gas density (or redshift) of the host cloud, which provides a lower boundary to the mass of the host cloud in which star formation is triggered by the first-generation supernova.
Also, the condition for fragmentation is very sensitive to 
the thermal state of interstellar gas.
Our result shows that for a reasonable range of temperatures ($200 {\rm K} \sim 1000$ K) of interstellar gas, the formation of second-generation stars can be triggered by a single supernova or hypernova with explosion energy in the above range, in a primordial cloud of total (the dark and baryonic) mass as low as $\hbox{a few} \times 10^{6} \msun$.  
For higher temperature in the interstellar gas, however, the condition for the fragmentation in the swept gas shell demands a larger supernova explosion energy.
  
We also follow the subsequent contraction of the fragment pieces assuming their geometry (sphere and cylinder), and demonstrate that the Jeans masses in the fragments decrease to well below a solar mass by the time the fragments become optically thick to the H$_2$ and HD lines.  
     The fragments are then expected to break up into dense cores whose masses are comparable to the Jeans masses and collapse to form low mass stars that can survive to date.  
     If the material in the gas shell is  mixed well with the ejecta of
 the supernova, the shell and low-mass stars thus formed are likely to
 have metals of abundance $\feoh \simeq - 3$ on average. 
This metallicity is consistent with those of the extremely metal-poor stars found in the Galactic halo. 
     Stars with such low metallicities of $\feoh < -5$ as HE0107-5240, recently discovered in the Galactic halo, are difficult to form by this mechanism, and must be produced in different situations.   

\end{abstract}


\section{Introduction} 
Star formation in the early Universe is believed to have played a critical role in the formation and evolution of galaxies.  
     Massive stars explode as supernovae that may contribute to the cosmic reionization and metal pollution of the Universe.
     Low-mass stars formed in the early Universe must survive to date, and are expected to carry precious information during, or even previous to, the early epoch of galaxy formation.  
     Because of this importance, the process of star formation in the early Universe has been attracting wide interest \citep[e.g.,][]{ostriker96,uehara96}.  

In the bottom-up scenarios of structure formation like those of the cold dark matter models, the first collapsed objects should have formed at redshifts of $z \sim 10^2 - 10$ with mass scales of $10^5 \msun - 10^8 \msun$ \cite[e.g., ][]{haiman96, tegmark97}.  
     Stars ought to have been born in these first collapsed primordial clouds, totally lacking  metals, before galaxies were formed.  
     Such stars are referred to as first-generation stars.  
     When the first collapsed objects are virialized, gas is once heated up to temperatures of 
$T \simeq 100 \, h^{2/3} \, (M/10^6 M_\odot)^{2/3} \, (1+z)$ K \citep{bromm02}, and hence, needs to be cooled efficiently to collapse further into stars in these primordial clouds.  
     Theoretical studies have suggested that in the first collapsed pregalactic objects, hydrogen molecules can form and cool the gas to temperatures of $T \simeq {\rm several} \times 10^2$ K, which leads to the formation of first-generation stars \citep[e.g.,][]{yoneyama72,pal83}.  
     First-generation stars are expected to be massive, or low-mass
     deficient due to temperatures higher than those of the present
     interstellar matter (\citealp{bromm99}; \citealp{bromm02}; \citealp{nakamura99}; \citealp{abel00}, but see \citealp{nakamura01}).

On the other hand, the number of extremely metal-poor stars found in our Galaxy has recently been increased substantially (more than 100 for stars with $\feoh < -3$) by large-scale surveys of the Galactic halo such as the HK survey (\citealp{beers92}\ and see also \citealp{norris99}) and Hamburg/ESO survey \citep{chri01}.  
     In particular, one giant star HE0107-5240 with a metallicity of $\feoh = -5.3$ has been found very recently \citep{chri02}.  
     For such low metallicities as $\feoh \lesssim -4$, metals contribute little to radiative cooling \citep{yoshii1980,omukai00}.  
     The thermal property of such metal-poor gas has to be essentially the same as that of the primordial gas.  
     The very existence of these low mass stars with such extremely low metallicities evidences the operation of the mechanism to form low-mass stars efficiently in the gas clouds of primordial abundances, completely devoid of metals.  
From existent studies on the thermal evolution of primordial gas clouds, the asserted lack in the low mass stars among the first-generation stars is mainly ascribed to the scantiness of relic electrons ($[{\rm e}/{\rm H}] \simeq  10^{-5}-10^{-6}$), which limits the formation of H$_2$ molecules necessary for  cooling of gas in the  primordial clouds \cite[e.g.][]{gal98, bromm02}.  
     The electron abundance can be larger, however, when gas in the primordial clouds is once heated above the temperatures of $\gtrsim 10^4$ K to be ionized and then undergoes cooling and recombination \citep{shapiro87}.   
     A larger fraction of free electrons will survive to yield  production of more abundant H$_2$ molecules primarily through ${\rm H}^-$ process of ${\rm H} + e^- \rightarrow {\rm H}^- + h\nu$ and ${\rm H} + {\rm H}^- \rightarrow {\rm H}_2 + e^-$.  
     Accordingly, the gas clouds can be cooled effectively first by H$_2$ molecules and then by HD molecules to temperature sufficiently low for the formation of low mass stars.  

There are two possible ways to achieve such situations in primordial clouds \citep[e.g.][]{shapiro87,ferrara98,nishi99,uehara00}.   
     For  primordial clouds more massive than $\sim 10^8 M_\odot$, the virialized temperatures at the collapse can be high enough to ionize the gas.   
     For such massive clouds, however, we have to take into account the possible influence of pollution, prior to the collapse, with metals produced by the first-generation stars, since their collapse is delayed on average for more than $\sim 10^8$ yr as compared with first collapsed clouds with masses of $\sim 10^6 M_\odot$ \citep{tegmark97}.   
     The other way may be shock heating by supernova explosions of the first-generation stars.   
     The ambient gas in the clouds must be swept together, heated up and ionized by the  shock waves induced by the supernova explosions.   
     
\citet{tsujimoto99} have proposed a scenario of  consecutive star formation, triggered by supernovae, from an abundance analysis of the extremely metal-poor stars found in the Galactic halo.  
     It is to be properly established, however, that a gas shell swept by a supernova shock can actually fragment to form stars within the host clouds.  
     \citet{nishi99} discussed the condition of cloud disruption by supernovae with hydrogen chemistry taken into account.  
     They argued that the scenario is possible only for massive first-collapsed objects with total masses larger than several $\times 10^7 \msun$.  
     Since they did not solve the thermal evolution of supernova remnants, however, the electron density, and hence, the resultant H$_2$ abundance may be underestimated, while the kinetic energy of supernova remnants is overestimated by applying the Sedov-Taylor solution to the later evolution.  
    \citet{ferrara98} discussed the notion that the gas, which is cooled to temperatures of $\sim 300$ K, is blown away from the cloud without fragmentation, in his study of the thermal evolution of a shocked gas,  taking into account the collective supernovae of the Population III objects of total masses from $10^6 \msun$ to $10^7  \msun$.  

Recently, \citet{sal03} have studied star formation induced
 by primordial supernova remnants (SNRs) in first collapsed objects,
  taking into account the H$_2$ cooling.  
They suggested that a supernova shock can trigger formation of low-mass
 stars only when it is driven by an
 efficiently energetic supernova of explosion energy greater than 
$10^{52}{\rm erg}$ 
 (e.g., hypernova or pair-instability supernova).
In their calculations, they assumed that the interstellar matter 
in which the SNR expands is ionized by UV radiation of the progenitor star 
and its temperature remains constant at $10^4$K during  expansion of 
the SNR.
The resulting high interstellar pressure tends to stall the expansion 
of the SNR shell, preventing the shell from sweeping and gathering enough gas to become self-gravitating.
However, the temperatures of interstellar gas are likely to 
decrease to lower values due to efficient radiative
 cooling by H$_2$ molecules that are reformed in the
 interstellar gas on a timescale of less than $10^3 \sim 10^4$ yr 
after the supernova explosion.
This reduction in the ambient pressure may significantly affect the
expansion of the SNR shell and changes the condition to trigger star formation.
\cite{bromm03} have also performed SPH simulations of such an SNR driven by 
a pair-instability supernova explosion due to a first-generation star.
They showed that a high-energy supernova explosion (pair instability supernova)
 blows a minihalo of mass  $10^6 \msun$
 and gas with metallicity $Z \gtrsim 10^{-2} Z_{\odot}$ is ejected to
 the intergalactic space. 
However, they did not address the issue of star formation triggered by the first supernova itself.

In this previous work \citep{ferrara98,sal03,bromm03},
 only H$_2$ molecules have been considered as
 the coolants after recombination.   
It should be pointed out, however, that under some conditions
 in the primordial clouds, deuterated hydrogen molecules, HD, 
 become a more efficient coolant than H$_2$ \citep[e.g., ][]{puy93, gal98}.  
For example, \citet{uehara00} have demonstrated
 that the HD cooling becomes dominant in the post-shock gas
 with shock velocities larger than $\simeq 300 {\rm km s}^{-1}$. 
\citet{flower02} and \citet{flower03} also investigated the effect of HD cooling on fragmentation of primordial clouds.    
\citet{nakamura02a} found that if H$_2$ abundance exceeds a critical value of $\sim 3 \times 10^{-3}$, the thermal evolution of primordial clouds is controlled by HD cooling, which reduces the gas temperature to $\sim 50-100$ K \citep[see also ][in more detail]{nakamura02b}.
     These studies indicate that  HD cooling plays an important role
 in star formation in primordial gas clouds.  

In the present paper, we investigate the dynamical and thermal evolution of a gas shell swept up by an SNR in a first collapsed primordial cloud and discuss the possibility of low-mass star formation that can survive to date.  
     We take into account the chemistry of H$_2$ and HD molecules, and
 the evolution of an SNR is studied using a semi-analytic formula, which
 enables us to sweep the large parameter space with detailed treatment
 of the primordial chemistry.  
 \citet{uehara00} have computed the evolution of a gas cloud for a particular initial condition, and demonstrated that once the gas has been ionized, the HD molecule plays a dominant role in cooling and enables low mass star formation. 
    Our purpose is then to explore the conditions of explosion energy and the redshift of  supernovae in primordial clouds that can lead to low mass star formation. 
     First, we study the condition whether the gas shell, swept by an SNR, can break up into fragments as a result of cooling due to H$_2$ and HD molecules. 
     Further, we follow the evolution of the fragments into lower mass cores from which low-mass stars will be formed. 
   In the computations, we also solve the thermal evolution of ambient gas simultaneously with  ionization by UV radiation from a progenitor star, and discuss the effects of the thermal state of ambient gas on the evolution of the SNR.  
     Since we are interested in low-mass star formation, triggered by a supernova explosion in a very first collapsed object, we work on the conditions appropriate to such situations.  
     The approximation and method of computations are given in \S 2, and the results of our computations will be presented in \S 3.  
     In \S 4, we derive the conditions under which supernova explosions of first-generation stars can trigger low-mass star formation in first collapsed clouds as functions of collapse redshift and energy of the explosion.
We discuss the relevance of our results in relation to the observed extremely metal-poor stars in \S 5.  

\section{Model}
A massive first-generation star explodes as a supernova, sweeping up the ambient medium to form an expanding gas shell.
If the gas shell becomes self-gravitating, it is likely to break up into fragments, which are expected to re-fragment into denser cores where the next-generation stars are formed. 
We explore the evolution of a supernova remnant (SNR) from the expansion of a shock wave in the ambient gas to the formation of dense cores in the fragments of the swept gas shell.
The evolution of the SNR depends on two parameters: the explosion energy of the supernova, $\varepsilon_0$, and the density of the interstellar gas, $\rho_0$, in the host cloud.  
For given parameters, we solve the expansion and variations in the structure of the SNR by applying a semi-analytical approximation to the blast wave expanding in the uniform density.  
The chemical compositions in the gas shell swept by the supernova shock, which play a critical role in determining the thermal state of the SNR, are calculated by directly integrating the rate equations.
To estimate the temperature and pressure in the ambient gas, we also solve the chemical and thermal evolution of the ambient gas.
We describe our methods and approximations in the following subsections.  

\subsection{Timescales}
\label{sec:tsl}
We start by defining four typical timescales that characterize the
evolution of an SNR, i.e., the expansion timescale $\tau_{\rm exp}$, the free-fall timescale $\tau_{\rm ff}$, the dynamical timescale $\tau_{\rm dyn}$, and the cooling timescale $\tau_{\rm cool}$. 

The expansion timescale is defined as
\begin{equation}
 \tau_{\rm exp}\equiv \frac{R}{\dot{R}} \, ,
 \label{eq:texp}
\end{equation}
 where $R$ and ${\dot{R}}$ denote the radius and the expansion speed of an SNR,
 respectively.
The cooling timescale is given by 
\begin{equation}
\tau_{\rm cool}\equiv\frac{n k  T  }{(\gamma-1)\  \ \Lambda(T, n, \rm{Compositions})} \, ,
\label{eq:cool}
\end{equation}
where $T$ and $n$ are the temperature and the number density in the swept gas shell, and $k$ and $\gamma$ are the Boltzmann constant and the ratio of specific heats, respectively.  
   The symbol $\Lambda$ (${\rm erg \ cm^{-3} s^{-1}}$) denotes the total energy loss rate per unit volume, which sums up all the cooling rates that are summarized in Appendix A. 

The dynamical timescale represents the time in which the sound wave crosses the gas shell as 
\begin{equation} \tau_{\rm dyn}\equiv\frac{\Delta R}{c_{\rm s}}
    =\frac{R \rho_0/3\rho}{\left({\gamma k T}/{\mu m_{\rm a}}\right)^{1/2}} \, ,
\label{eqn:tdyn}
\end{equation}
where  $\rho_0$ is the density of the ambient gas; 
     $\rho$, $c_{\rm s}$, $\Delta R$ and $\mu$ denote the gas density, the sound speed in the shell, the shell width, and the mean molecular weight of the shell, respectively: 
     $m_{\rm a}$ is the atomic mass unit ($\rho = \mu m_{\rm a} n$).

The free-fall timescale is written as
\begin{equation}
 \tau_{\rm ff} \equiv \left( \frac{C}{G \rho}   \right)^{1/2},
\label{eqn:tffs}
\end{equation} 
     where $G$ is the gravitational constant and $C$ is a structure parameter, given by $C=3\pi/32$ and $C=8/\pi$ for the spherical and cylindrical collapses, respectively.  
     Note that the free-fall timescale for the cylindrical collapse ($\tau_{\rm ff,c}$) is larger than that for the spherical collapse ($\tau_{\rm ff,s}$) by a factor of 2.94.  
See \S 2.2 for more detail.

\subsection{Expansion of SNR and Star Formation in the Fragments}
In our calculation, we treat the entire evolution from expansion of the supernova remnant shell,  fragmentation of the swept gas shell and contraction of fragments in the expanding shell, through the formation of dense core consistently.  
     In the following, we divide the evolution into two phases, `{\it the ante-fragmentation phase}' and `{\it the post-fragmentation phase}' for convenience's sake.  
     The ante-fragmentation phase is defined as a period in which a supernova shock propagates and gathers interstellar gas until the swept gas shell is affected by gravitational instability and undergoes fragmentation.  
     The post-fragmentation phase is defined as a period after fragmentation occurs until the fragments contract to be optically thick against the line
 emissions by H$_2$ and HD molecules.  

\subsubsection{The ante-fragmentation phase}
Evolution of SNRs was studied extensively in the early 1970's using one-dimensional hydrodynamic codes  \citep{cox72,chev74}.   
     Summarizing their results, the evolution of SNRs is divided into three stages as (1) {\it the free-expansion stage}, (2) {\it the Sedov-Taylor adiabatic stage}, and (3) {\it the pressure-driven expansion stage} \citep[see also ][]{ost88}.  
     As shown below, the expanding gas shell comes to be dominated by self-gravity and is expected to fragment during the pressure-driven expansion stage. 

\noindent (1) {\sl the free-expansion stage} 
($\tau_{\rm exp} \ll \tau_{\rm cool}$)\hfill\break
     In this stage, the SN ejecta expand freely.   
     After the ambient gas swept by an SNR surpasses the SN ejecta in mass, the expansion is decelerated and the SNR enters the Sedov-Taylor adiabatic stage. 
     With the ambient density $\rho_0$ and the ejecta mass $M_{\rm ej}$, the transition radius is given by 
\begin{equation} 
R_1= (3M_{\rm ej}/4\pi\rho_0)^{1/3}. 
\label{fe2stradius}
\end{equation} 
     We put $M_{\rm ej} = 10 \msun$, for simplicity. 
     We have confirmed that the choice of the ejecta mass hardly affects the later evolution.  

From comparison with the Sedov-Taylor self-similar solution (see eq. [\ref{eqn:exp-adiabatic}] below), the transition time from the free-expansion to Sedov-Taylor stages is given as 
\begin{equation}
t_1 = \left[ \frac{R_1}{1.15} \left( \frac{\rho_0}{\varepsilon_0} \right)^{1/5} \right]^{5/2}
 = \left[ \frac{1}{1.15} \left(\frac{3M_{\rm ej}}{4\pi\rho_0}\right)^{1/3}  \left( \frac{\rho_0}{\varepsilon_0} \right)^{1/5} \right]^{5/2},
\label{eq:freet}
\end{equation} 
     where $\varepsilon_0$ represents the SN explosion energy. 

\noindent (2) {\sl The Sedov-Taylor adiabatic stage} ($\tau_{\rm exp} < \tau_{\rm cool}$) \hfill\break
After the transition time $t_1$, the SNR is dominated by the blast wave, and hence, the structure tends to a self-similar solution \citep{sedov46,taylor50}.  
     The expansion of the shock front is well approximated to the Sedov-Taylor solution and described as 
\begin{equation}
R =1.15 \left(\frac{\varepsilon_0}{\rho_0}\right)^{1/5}t^{2/5},
\label{eqn:exp-adiabatic}
\end{equation}   
and the expansion speed of the SNR shock front is expressed by
\begin{equation}
V = \frac{dR}{dt} = 0.46 \left(\frac{\varepsilon_0}{\rho_0}\right)^{1/5} t^{-3/5}.
\label{eqn;exp-vel}
\end{equation}
This gives the post-shock pressure and shock temperature as 
\begin{eqnarray}
P_{\rm p s} & = & 0.42 \frac{\rho_0}{(\gamma+1)}\left( \frac{\varepsilon_0}{\rho_0} \right)^{2/5} t^{-6/5}, \label{eqn:pst} \\
T_{\rm p s} & = & \left( \frac{8}{25} \frac{\mu m_{\rm a}}{k} \right)  
\left( \frac{\varepsilon_0}{\rho_0}\right)^{2/5}t^{-6/5}.
\label{eqn:T-sedov}
\end{eqnarray}  
     These equations are derived from the Rankine-Hugoniot relation and the Sedov-Taylor solution, and indicate that the post-shock pressure and temperature decrease with the expansion of SNR 
(\citealt{ost88}; see also \citealt{saka96}).  
     The post-shock gas is also cooled via radiative energy losses. 
     Equation (\ref{eq:cool}) enables us to estimate the cooling time just behind the shock front. 
     After the cooling time-scale becomes shorter than the expansion time-scale, i.e. $\tau_{\rm cool}<\tau_{\rm exp}$, a cooled shell will form just inside the shock front and the SNR enters  the pressure-driven expansion stage.  
     We denote the time when $\tau_{\rm cool}$ decreases to be equal to $\tau_{\rm exp}$ by $t_2$.

\noindent (3) {\sl The pressure-driven expansion stage} ($\tau_{\rm cool} < \tau_{\rm exp}$ and $\tau_{\rm dyn} < \tau_{\rm ff}$) \hfill\break 
The shell expansion is driven by the high pressure in the hot low-density cavity. 
The equation of motion of the shell is written as
\begin{equation}
 \frac{4\pi \rho_0}{3}\frac{d R^3 \dot{R}}{dt}=4\pi R^2
\left( P_{\rm in} - P_{\rm hc} \right),
\label{eqn:exp-pd}
\end{equation} 
where $P_{\rm in }$ and $P_{\rm hc}$ mean the pressure in the inner
cavity and the ambient gas pressure \citep{saka96}, respectively.
We assumed that the pressure inside the cavity decreases adiabatically as   
\begin{equation}
P_{\rm in} = P_2 \left( \frac{R}{R_2} \right)^{-3 \gamma} . 
\label{eq:pin}
\end{equation}
In this equation, $R_2$ and $P_ 2$ represent the radius of the shock front and the post-shock pressure, respectively, which  are given from equations (\ref{eqn:exp-adiabatic}) and (\ref{eqn:pst}) at the beginning of the pressure-driven expansion stage ($t = t_2$).   
The ambient gas pressure $P_{\rm hc}$ is derived by calculation of the chemical reaction and thermal evolution of ambient gas, as shown in \S 2.4.

Using the above equation (\ref{eqn:exp-pd}), we calculate the SNR expansion until fragmentation occurs in the shell (see below) or the expansion velocity of the SNR shell, $V$, becomes slower than the sound speed in the ambient gas of the host cloud, $c_{\rm s, hc}$.   
As a result of the radiative cooling, the free-fall time scale $\tau_{\rm ff}$ grows shorter than the sound-crossing time ($\tau_{\rm dyn} = \Delta R/c_s$) over the shell width $\Delta R$, and the self-gravity finally becomes dominant over the pressure force in the shell.  
     Accordingly, we may well assume that the shell breaks into spherical fragments or cylindrical filaments when the following condition is satisfied: 
\begin{equation}
\tau_{\rm ff}=\tau_{\rm dyn}, \quad \hbox { or } \quad \Delta R = c_s (C/G \rho)^{1/2}.  
\label{eq:fragcond}
\end{equation}
     The shell width $\Delta R$ is related to the shell density $\rho$ as $4\pi \Delta R R^2 \rho=(4\pi/3)R^3 \rho_0$, where we used the mass conservation relation.  
     Note that this gives $\Delta R=R/12$ for the Sedov adiabatic stage if $\gamma=5/3$.  
     As a corollary, the above condition $\tau_{\rm ff}=\tau_{\rm dyn}$ is equivalent to the condition that the Jeans length in the shell is equal to the shell thickness. 
     Such fragments produced by the break-up of the gas shell may correspond to the filamentary structures seen in the SNR, such as the Cygnus loop. 
 \citet{bromm03} showed that the evolutions of SNRs
 of first-generation stars are well approximated
 by the analytical solutions used in the present paper.  
However, due to the restriction of numerical simulations
 (limited spatial resolutions) they could not explore
 the process after fragmentation.  
We solve the contraction of the fragmented pieces
 with a semi-analytical method shown below and figure out the minimum mass of the second-generation stars.

\subsubsection{The post-fragmentation phase} 
After the gas shell fragments, each fragment piece begins to contract in a free-fall timescale. 
     In this phase, we consider two different geometries for the fragment: the spherical and cylindrical configurations.  
     For the spherical case, the contraction obeys the following equation:
\begin{equation}
\frac{dv}{dt} = - \frac{G m_r}{r^2},
\end{equation}
where $v$ and $m_r$ represent the infall velocity and the mass contained 
 within the radius $r$.
     For a uniform sphere, the timescale of contraction agrees with the free-fall timescale expressed by equation (\ref{eqn:tffs}).
     Since the SNR shell fragments due to its self-gravity, on the other hand, the fragment is likely to have cylindrical geometry and the gas contracts in the radial direction, i.e., perpendicularly to the cylinder axis.  
      Virial analysis of the cylindrical isothermal filament gives the equation of motion for the filament as follows:
\begin{equation}
\frac{dv}{dt}=-\frac{2G}{r} \left( \lambda - \lambda_c  \right)
\label{eq:mec}
\end{equation}
\citep{ostriker64,uehara96}, where $\lambda$ denotes the line mass of the filament [= $\pi(\Delta R/2)^2\rho$], and  $\lambda_c$ is the critical value corresponding to the line mass of the filament in hydrostatic equilibrium and given by
\begin{equation}
\lambda_c = \frac{2kT}{\mu m_{\rm a} G} =\frac{2c_s^2}{G}.
\end{equation}
     This equation indicates that the line mass density must be larger than $\lambda_c$ for the cylindrical fragment to contract.  This condition is rewritten as 
\begin{equation}
\tau_{\rm dyn} > \sqrt{\frac{8}{\pi G \rho}} \equiv \tau_{\rm ff,c}
\end{equation} 
     in terms of the free-fall timescale for a cylinder with uniform density, $\tau_{\rm ff,c} =(8/\pi G \rho)^{1/2}$ from equation (\ref{eqn:tffs}).  

As shown in the next section, the main coolants in the fragments are H$_2$ and HD line cooling. 
     If the fragments are optically thick to these line emissions, the subsequent contraction proceeds nearly adiabatically, and along with the rise in the temperature, the right-hand side of equation (\ref{eq:mec}) becomes no longer negative.   
     Then, the radial contraction of the fragment will be terminated, and the fragmentation is expected to occur again along the cylinder axis. 
     Accordingly,  we terminate following the contraction of the fragment when the fragments become optically thick to the line emissions of the most efficient coolant, either HD or H$_2$, where the optical depth is evaluated under the escape probability method. 
     In order to estimate the optical depths of H$_2$ and HD lines, we use the Large Velocity Gradient (LVG) method \citep[e.g.,][]{gold74}.  
     The optical depth for a transition $J+1 \rightarrow J$ is given by 
\begin{equation}
\tau _{J+1,J} = \frac{hc}{4\pi}\frac{B_{J,J+1} n_J}{|dv/dr|}
\left( 1-\frac{g_J n_{J+1}}{g_{J+1} n_J} \right),
\end{equation}
where $h$ is  the Plank constant, $c$ is the speed of light, $B_{J,J+1}$ is the Einstein's $B$ coefficient, and $n_J$ and $g_J$ are the number density and statistical weight  of the $J$-th level, respectively. 
The number density of the $J$-th level is evaluated  using a two-level transition model with fitting formulas of the collisional de-excitation rates by \citet{gal98} and \citet{flower99}.
     The first six rotational transition levels are taken into account.
The velocity gradient, $|dv/dr|$, is approximated as $\alpha v_{\rm th}/R_J = \alpha (\pi G \rho)^{1/2}$, where $\alpha$, $v_{\rm th}$, and $R_J$ are non-dimensional numerical constant, the thermal velocity of HD, and the filament radius, respectively.
The value of $\alpha$ depends on the velocity distribution of the collapsing filament. 
When the density reaches a critical density of HD cooling, the contraction of the filament becomes quasi-static and therefore, the velocity gradient becomes subsonic \citep{nakamura02b}.
We thus set $\alpha$ to unity, for simplicity.

It is thought that the Jeans mass, when the fragment becomes optically thick against the line emissions of the most efficient coolant (H$_2$ or HD), gives the characteristic minimum mass for stars that will be formed in the fragment pieces, and hence, if it is much smaller than $\sim 0.8 \msun$, we may well expect that the low mass stars surviving today can be formed.  
For respective stages, we study the thermal and chemical histories of the SNR shell by a one-zone approximation, which is explained below.

\subsection{Temperature and Density in the Shell}
\subsubsection{The ante-fragmentation phase} 
The temperature at the transition epoch, $t_1$, from the free-expansion to the Sedov-Taylor adiabatic stage, is determined from the condition that the transition is continuous as
\begin{equation}
T(t_1) = (1.15)^5 {\pi \over 25} {\mu m_{\rm a} \over k} {\varepsilon_0 \over M_{\rm ej}}
\label{eqn:initemp}
\end{equation}
 with $\gamma = 5/3$.
     We calculate the variation in the temperature of the shell under the one-zone approximation in the following way:
\begin{equation}
\frac{dT}{dt}=\left(\frac{dT}{dt}\right)_{\rm exp} 
              +\left(\frac{dT}{dt}\right)_{\rm rad}
              +\left(\frac{dT}{dt}\right)_{\rm comp},
\label{eqn:dTdt-collapse}
\end{equation}
where $(dT/dt)_{\rm exp}$, $(dT/dt)_{\rm rad}$, and $(dT/dt)_{\rm comp}$ represent the term due to the expansion cooling, the radiative cooling, and the compressional heating due to contraction of the shell, respectively.  
     In the Sedov-Taylor adiabatic stage, the postshock temperature is given by the similarity solution, where $T \propto t^{-6/5}$.  
     Thus, the expansion cooling term is denoted as
\begin{equation}
\left( \frac{dT}{dt} \right)_{\rm exp} = -1.2 \ \frac{T}{t}.
\end{equation}
This term is only effective in the Sedov-Taylor adiabatic stage, and in the pressure-driven expansion stage, tends to be negligible as compared with the radiative cooling and the compressional heating term. 
Thus, we include the expansion cooling only in the Sedov-Taylor stage and  ignore it after the pressure-driven stage. 

The radiative cooling term is denoted as 
\begin{equation}
\left( \frac{dT}{dt} \right)_{\rm rad}= -{ T \over \tau_{\rm cool}} = - T \frac{(\gamma-1) \ \Lambda(T, n, {\rm Composition})}{P} .
\end{equation} 
     In this cooling, we include the inverse Compton cooling \citep{ikeuchi86} and the radiative cooling by the atoms and ions of H and He \citep{cen92} and by H$_2$ and HD molecules \citep{gal98,flower00}.  
     We take into account the effect of the CMB radiation on the radiative cooling by modifying the total cooling rate as 
\begin{equation}
\Lambda = \Lambda(T) - \Lambda(T_{\rm CMB}) ,
\end{equation}   
     where the cosmic background temperature $T_{\rm CMB}$ is calculated as $T_{\rm CMB}=2.73(1+z)$ K.
Owing to this formula, the gas cannot cool below the CMB temperature at that time.

The compressional heating has a form   
\begin{equation}
\left(\frac{dT}{dt}\right)_{\rm ad}=(\gamma-1)\frac{T}{n} \frac{dn}{dt},
\end{equation}
where $n$ represents the number density of molecules, atoms, ions
 and electrons.
This term becomes effective after the shell has cooled
 and the density increases.

The density in the shell is given by the jump condition across the strong shock front during the Sedov-Taylor stage  as 
\begin{equation}
     \rho = \frac{\gamma+1}{\gamma-1} \rho_0, 
\end{equation}
where $\rho_0$ represents the ambient gas density.  
     This equation indicates that the density in the shell is equal to $4\rho_0$ for $\gamma=5/3$. 
After the condition $\tau_{\rm cool} < \tau_{\rm exp}$ is realized
 in the pressure-driven stage,
 time evolution of physical quantities in the shell differs
 according to whether the cooling time is shorter than the sound crossing time
 of the shell ($\tau_{\rm cool} < \tau_{\rm dyn}$)
 or the other way around ($\tau_{\rm dyn} < \tau_{\rm cool}$).  
     In the former case, the shell cools without changing the width of the shell. 
Thus, we assume
\begin{equation}
     \frac{d\rho}{dt}=0, \qquad \qquad {\rm for}\ \tau_{\rm cool} < \tau_{\rm dyn}.
\end{equation}
     Accordingly, the pressure in the shell varies in proportion to the temperature given by equation (\ref{eqn:dTdt-collapse}) in this regime.  

If the sound-crossing time is shorter than the cooling time ($\tau_{\rm dyn}<\tau_{\rm cool}$), on the contrary, the shell is subject to compression, and in the decelerating frame with the gas shell, the structure tends to be in hydrostatic equilibrium with the boundary pressures given by the inner cavity pressure (eq.[\ref{eq:pin}]) and the post-shock pressure, $P_{\rm out}$.  
     Thus, we assume that the pressure in the shell is brought near to the average of $P_{\rm in}$ and $P_{\rm out}$ in a dynamical timescale as follows:  
\begin{equation}
     \frac{d P}{d t} =  \left( \frac{ P_{\rm in} + P_{\rm out}}{2} - P \right) \frac{1}{\tau_{\rm dyn}}, \qquad \qquad  {\rm for}\ \tau_{\rm dyn} < \tau_{\rm cool},  
\label{eq:dyncool}
\end{equation}
where the post-shock pressure is assumed as the sum of the ram pressure
 and the pressure in the ambient gas $P_{\rm hc}$: 
\begin{equation}
     P_{\rm out} = \frac{2}{\gamma +1} \rho_0 \dot{R}^2 + P_{\rm hc}.
\label{eq:pout}
\end{equation}
     The density variation at this stage is determined by equations (\ref{eqn:dTdt-collapse}) and (\ref{eq:dyncool}) with the equation of state.

\subsubsection{The post-fragmentation phase} 
During the pressure-driven expansion stage, the dynamical timescale lengthens as the shell expands, eventually reaching the free-fall timescale. 
     When the condition $\tau_{\rm ff} < \tau_{\rm dyn}$ is satisfied, the shell begins to fragment owing to  self-gravity, and the evolution enters the final stage of the fragmentation and star formation stage. 
     As mentioned above, the subsequent evolution of the fragments seems to differ according to their geometry.  
     We consider two different types of fragment geometry, i.e., the spherical and the cylindrical symmetric configurations. 
     For spherical fragments, the time variation of the density is approximated as 
\begin{equation}
     \frac{d\rho}{dt} = \frac{\rho}{\tau_{\rm ff,s}}
\label{eqn:dndt-sph}
\end{equation}
     with the free-fall timescale given by equation (\ref{eqn:tffs}).

On the other hand, if the fragments form long cylindrical shapes, the widths of the filaments shrink with time according to equation (\ref{eq:mec}).  
    Since equation (\ref{eq:mec}) gives the contraction time scale of $\simeq r/[2G(\lambda-\lambda_c)]^{1/2} \simeq 1/[2\pi G \rho (1-\lambda_c/\lambda)]^{1/2}$, we assume that the density changes as
\begin{equation}
     \frac{d\rho}{dt} = \frac{\rho}{\tau_{\rm ff,c}(\rho)} \left(1 - \frac{ \lambda_c }{\lambda} \right)^{1/2}, \qquad (\lambda > \lambda_c) 
\label{eqn:dndt-cyl}
\end{equation}
     with the line mass density $\lambda$ determined at the fragmentation \citep{ostriker64}. 

The cylindrical collapse proceeds much slower than that of the spherical one \citep{uehara96}.
     This comes from the factor $1-\lambda_c/\lambda$ in equations (\ref{eq:mec}) and (\ref{eqn:dndt-cyl}).
Along with the cooling of the gas, $\lambda_c$ decreases.
If the critical line mass $\lambda_c$ becomes smaller than the line mass, i.e., $\lambda > \lambda_{c}$, the filament contracts.  
     At the same time, the contraction causes $\lambda_c$ to increase because of the compressional heating. 
If $\lambda_c$ increases to approach $\lambda$,
 the contraction slows down,
 and finally halts  when $\lambda_c$ reaches $\lambda$. 
Accordingly the contraction of filaments proceeds in the cooling timescale,  keeping the temperature slightly lower than determined from the condition $\lambda_c = \lambda$, and hence, much more slowly than in the free-fall timescale.  
     This makes a clear contrast to the fast collapse accomplished for the spherical system [eq.(\ref{eqn:dndt-sph})]. 

In our calculation, the initial line mass $\lambda$ is
 determined at the fragmentation epoch ($\tau_{\rm ff} = \tau_{\rm dyn}$)
 under the assumption that fragmentation occurs with a wavelength
 equal to the shell width.  
This gives a line mass of 
\begin{equation}
     \lambda = \pi \left(\frac{\Delta R}{2}\right)^2 \rho=\frac{\pi R^2 \rho_0^2}{36 \rho},
\end{equation}
 where we used the shell thickness of $\Delta R=\rho_0 R/3\rho$. 
As the shell expands, the fragments accumulate additional gas that is swept 
 by the shell.  
We assume that the line mass increases in proportion to the shell mass as
\begin{equation}
\frac{d\lambda}{dt}=\lambda \frac{\dot{M}_{\rm sw}}{M_{\rm sw}},
\end{equation}   
 where $M_{\rm sw}$ and $\dot{M}_{\rm sw}$ represent the shell mass and
 the rate of increases in the shell mass, respectively.
Along with the increase in $\lambda$, the contraction becomes shorter than the expansion, and thereafter, $\lambda$ stays constant, since the contraction is further accelerated with the decrease in the free-fall and cooling timescales with the increases in the density.    
We also assume
 that the mass of the spherical fragments increase in proportion to the 
 shell mass.  
However, since the spherical collapse is much faster than 
 the cylindrical one, this does not play an important role.

\subsection{The evolution of the ambient gas}

The thermal state of the ambient gas may play a part in the evolution of
 the expanding gas shell in the pressure-driven expansion phase (\S 2.2.1 (3))
 and the resultant fragmentation of the swept shell.  
In the course of formation of first-generation stars, gas in the primordial
 clouds suffers cooling by H$_2$ molecules formed with relic electrons as agents, and the   gas temperature can 
decrease as low as $\sim 200$ K \citep[e.g., ][]{bromm02}.
Once a massive star is born, however, gas is ionized by the radiation of the massive   progenitor stars and heated to 
$T\sim 10^4{\rm K}$. 
After the massive star explodes, radiation coming from the interior of the SNR may heat  the  ambient gas or destroy the hydrogen molecules.

As for the UV radiation from a progenitor star,
the extent of the ionized region depends on the ratio of the total number of ionization  photons per unit time to the 
recombination rate of the hydrogen.
If we assume an O5-type progenitor star and the ambient gas density of
 $n_{0} =1$ cm$^{-3}$, this star ionizes the ambient gas inside the
 Str\"{o}ngren radius of $\simeq 100$ pc (Panagia 1973).
After the massive star explodes, the ambient ionized gas begins to recombine.
Although the ambient gas continues to cool by radiative cooling,
 the gas near the shock front (preshock gas) may be heated by the ionization
 radiation from the SNR's hot interior (Shelton 1999, Slavin \etal 2000).

The preshock gas ionized by the radiation emitted
 from the postshock hot gas is calculated 
 by Shull \& McKee (1979) for a metal-rich gas cloud
 and by Shull \& Silk (1979) for a metal-poor gas cloud.
The controlling parameter for this problem
 is the ratio of the emergent photon flux $\Phi({\rm cm^{-2}s^{-1}})$
 from the postshock gas
 to the hydrogen atom number flux $n_{0} V ({\rm cm^{-2}s^{-1}})$ 
 flowing into the SNR.
Shull \& Silk (1979) showed
 that the emitted photons scarcely ionize the ambient gas for $n_{0}=1$ cm$^{-3}$
 if the shock velocity is as low as $V  \lesssim 60$ km\,s$^{-1}$,
 because the above photon-to-gas ratio $\Phi/n_{0} V \ll 1$ is too small
 at that time. 
Thus, the heating of the preshock gas by the SNR shock is negligible in
 the late evolutionary phase ($V  \lesssim 60$km s$^{-1}$). 
In the early evolutionary phase, the emission from the SNR shock is
 also negligible, because the shock is strong and the ambient gas pressure
 does not play a role, as discussed in Shapiro \& Kang (1987). 

In our calculation, as long as the shock speed is faster than 
 $V  > 60 \, {\rm km s}^{-1}$,
 the ionization level of the ambient gas is always much higher than that expected from
 the ionization by the SNR (Shull \& McKee 1979) and its temperature is maintained at 
 $T_{\rm hc} \sim 10^4$K,
 because the ambient gas photoionized by the progenitor star has not 
 recombined/cooled sufficiently by this epoch.
Thus, we expect that SNR evolution is not affected, whether the ionizing photons
 emitted from the SNR are considered or not.
Therefore, we neglect  ionization by the ionization photons from the SNR.

The SNR radiation may prevent the ambient gas from cooling
 through the dissociation of the H$_2$ molecules, which is the effective coolant at lower    temperature ($T<10^4$K).
The dissociation timescale of the molecular hydrogen, $t_{\rm dis}$, is given by
 $t_{\rm dis} = 2.8 \times 10^{-16}\, F_{\rm LW}^{-1}$\ yr,  (Omukai \& Nishi 1999). 
Here $F_{\rm LW}$ (erg s$^{-1}$ cm$^{-2}$ Hz$^{-1}$) is the average radiation flux
 in the  Lyman - Werner (LW) bands.
The flux $F_{\rm LW}$ is estimated as a function of the shock velocity of SNR (Shull \&  McKee 1979; Shull \& Silk 1979).
The hydrogen molecules begin to form after the shock velocity decelerates
 below $V = 60$ km s$^{-1}$ in our calculation.
From the flux $F_{\rm LW} \simeq 1.3 \times 10^{-22}$ erg s$^{-1}$ cm$^{-2}$ Hz$^{-1}$ for $n_{0} = 1$ cm$^{-3}$ at the stage of $V = 60$ km s$^{-1}$ (Shull \& Silk 1979),
 the dissociation timescale is given as $t_{\rm dis} = 2.2 \times 10^6$ yr,
 which is much longer than the formation timescale of the molecule hydrogen, 
 $n_{\rm H_2} /(dn_{\rm H_2}/dt)$.
After that, the dissociation timescale continues to increase, because the flux 
 $F_{\rm LW}$ decreases with time, and keeps longer than the formation timescale.
Therefore,  radiation from an SNR barely decreases the molecular fraction
 of ambient gas.

As a result, we can safely ignore the effect of  radiation from hot gas in the SNR
 and include only the effects of photoionization and heating due to the progenitor star.
However, the above effect may delay  cooling of the ambient gas slightly.
In \S 4, we discuss how the expansion of the shell is affected
 if we assume a higher-temperature ambient gas than calculated in the next section \S 3,
 and show the condition for  low mass star formation. 

In order to obtain the proper boundary conditions of SNR shell evolution,
 we need to know the evolution of the ambient pressure.
For this reason, we solve the chemical reactions and the thermal evolution
 of the ambient gas, simultaneously with the dynamical and thermal evolution
 of the SNR shell. 
The initial chemical composition of the ambient gas is derived
 under the assumption that the ambient gas has been heated and is kept
 at a temperature of $T_{\rm hc} = 10^4$ K by the progenitor star
 of the first generation which lives for $10^6$ yr. 
After the progenitor star explodes, we assume that the heating source
 disappears and follow the cooling of the ambient gas by solving
 the same rate equations and the equation of energy conservation,
 as described above, under the constant density of $\rho_0$. 
From the ambient temperature, $T_{\rm hc}$,
 we derive the interstellar pressure $P_{\rm hc}$ by
\begin{equation}
P_{\rm hc} = \frac{\rho_0\, k}{\mu_{\rm hc} \,m_a} T_{\rm hc},
\label{eq:phc}
\end{equation}
where $\mu_{\rm hc}$ is the mean molecular weight of the ambient gas. 
This pressure is applied to the evolution of the SNR gas shell in eq.\ (11). 
The SNR shell can continue expanding and sweeping the ambient gas, only when 
 the expansion speed is faster than the sound speed as
\begin{equation}
V  > c_{\rm s, hc}.
\label{cd:velexp}
\end{equation}
Otherwise ($V < c_{\rm s, hc})$,
 the gas shell dissolves and merges into the ambient gas.  
Accordingly, the above condition (eq.[\ref{cd:velexp}]) has to be satisfied
 for the gas shell to fragment.

\subsection{Chemical Composition}
We have modeled the thermal and dynamical evolution of the SNR shell and ambient medium under the one-zone approximation, as stated above. 
The cooling rates in the gas shell and ambient medium are determined by their chemical composition.
To estimate the abundances of  chemical species in these gases, we solve the time-dependent chemical reaction equations numerically. 
In this paper, we consider the chemical reactions of the following 12 species, H, H$^+$, H$^-$, He, He$^+$, He$^{++}$, H$_2$, D, D$^+$, HD, HD$^+$, and e$^{-}$. 
We adopt the primordial composition obtained by \citet{gal98} as the initial condition of our calculations. 
The reaction rates we include in our calculations are summarized in Appendix B.

\section{Results}
We begin our computation from the transition epoch, $t_1$, between the free-expansion and the Sedov-Taylor adiabatic stages given by equation (\ref{eq:freet}) for a given set of values of the explosion energy, $\varepsilon_0$, and the density, $\rho_0$, of the host cloud. 
     Then, we solve the equations of structural changes in the expanding SNR and the rate equations for the changes in the chemical abundances, formulated in the preceding section.  
     In the present work, we adopt an explosion energy between $\varepsilon_0 = 10^{51}{\rm erg}$ and $10^{52} {\rm erg}$, ranging from a normal supernova to  a hypernova \citep{nomoto99}.   The density of interstellar gas in the host cloud is taken to be in the range between $\rho_0 = 3.51 \times 10^{-25}{\rm g\,cm^{-3}}$ and $3.49 \times 10^{-23}{\rm g\, cm}^{-3}$  taking into account the formation epoch of the host clouds.
Since the densities in the virialized host clouds are approximated to 200 times
 the average baryon density in the Universe at that epoch \citep{white93},
 the above density range corresponds to the redshifts of formation epoch
 $z = 10 - 50 \ (\rho_0 \propto [1+z]^3)$ for the Einstein-de-Sitter Universe
 model with a Hubble constant of
 $H_0 = 70 \hbox{ km s}^{-1} \hbox{ Mpc}^{-1}$
 and the baryon fraction of $\Omega_{\rm B} = 0.06$.  

The model parameters adopted are listed in Table~\ref{table:results} with the characteristic physical quantities of gas shells and ambient gas when the condition of fragmentation, $\tau_{\rm ff} = \tau_{\rm dyn}$, is fulfilled.  
     In the following, we discuss the results, dividing the evolution into two phases, defined in the preceding section; {\it the ante-fragmentation phase} ($\tau_{\rm ff} > \tau_{\rm dyn}$) until the gas shell undergoes fragmentation and {\it the post-fragmentation
phase} ($\tau_{\rm ff} < \tau_{\rm dyn}$) during which the fragments collapse gravitationally, eventually forming second-generation stars. 

\subsection{The Ante-Fragmentation Phase}

\begin{figure}
\plotone{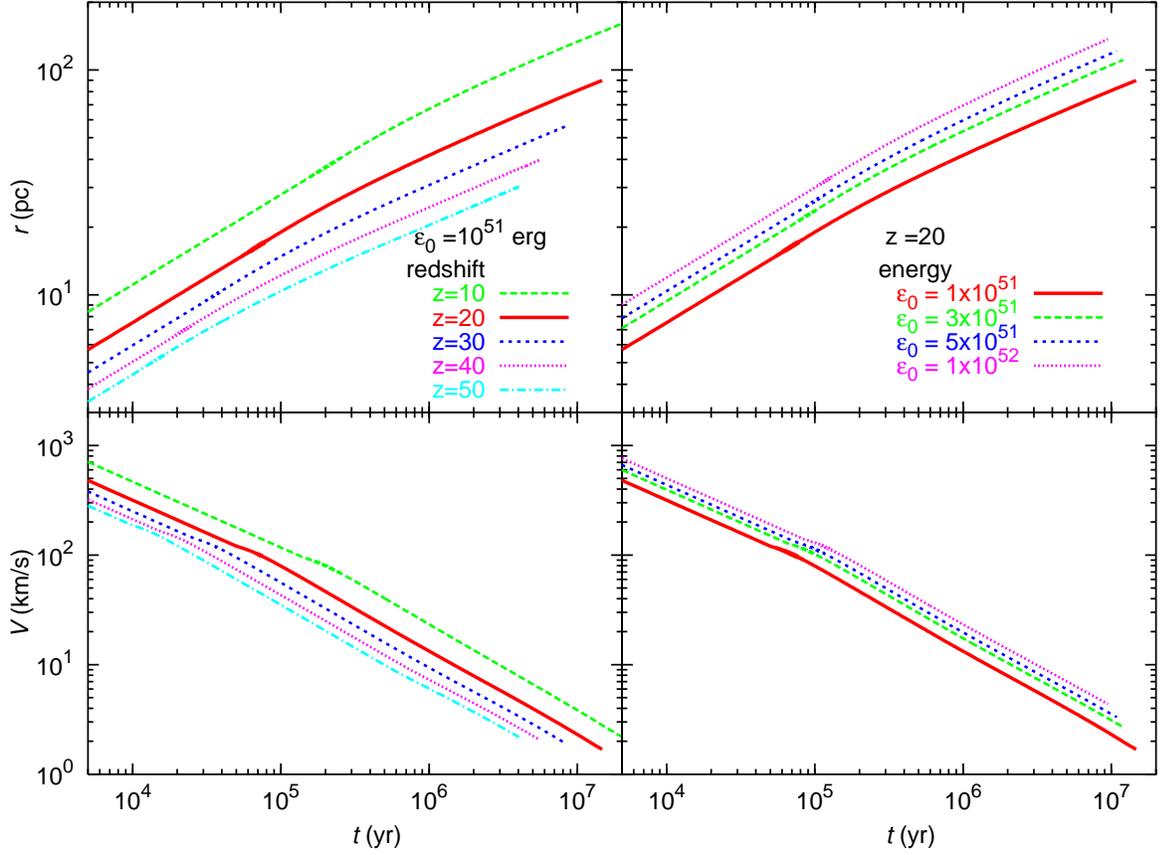}
\caption{
Evolution of SNR radius (top panels) and the shell expansion velocity (bottom panels) is compared among the models of different ambient densities ($z$=10, 20, 30, 40, and 50) but a fixed explosion energy ($\varepsilon_0 =10^{51}$ erg; left panels) and among the models of different explosion energies ($\varepsilon_0= 1, 3, 5, \hbox{ and} 10 \times 10^{51}$erg) but a fixed ambient density ($z = 20$; right panels).  
}
\label{fig:1}
\end{figure}

Figure~\ref{fig:1} shows the evolution of the shell radius and the
     expansion velocity of the SNR against the elapsed time from the SN
     explosion for various models. 
     For each model, the expansion of the SNR is well expressed by two power-laws, in which the break occurs at the transition time $t_2$ from the Sedov-Taylor ($R \propto t^{2/5}$; eq.~[\ref{eqn:exp-adiabatic}]) to the pressure-driven stages  ($R \propto t^{2/7}$; eq.~[\ref{eqn:exp-pd}]). 
The expansion velocity, and hence, the radius of the SNR, are 
larger for smaller collapse redshifts and for larger explosion energy.
For example, the initial radius at the transition epoch $t_1 = 2.1 \times 10^3$yr for $z = 10$ is 4.7 times as large as that at $t_1 = 2.6 \times 10^2$yr for $z = 50$. 
We note that the expansion velocity and radius depend weakly on ambient 
density, as inferred from the Sedov-Taylor similarity solution
[$ R \propto (\varepsilon_0/\rho_0)^{1/5}$], but the ambient density depends strongly 
on the collapse redshift [$\rho \propto (1+z)^2$]. 
Therefore, the expansion velocity and radius change greatly with $z$.  
More quantitative results of all models are summarized in Table \ref{table:results}.

\begin{figure}
\plotone{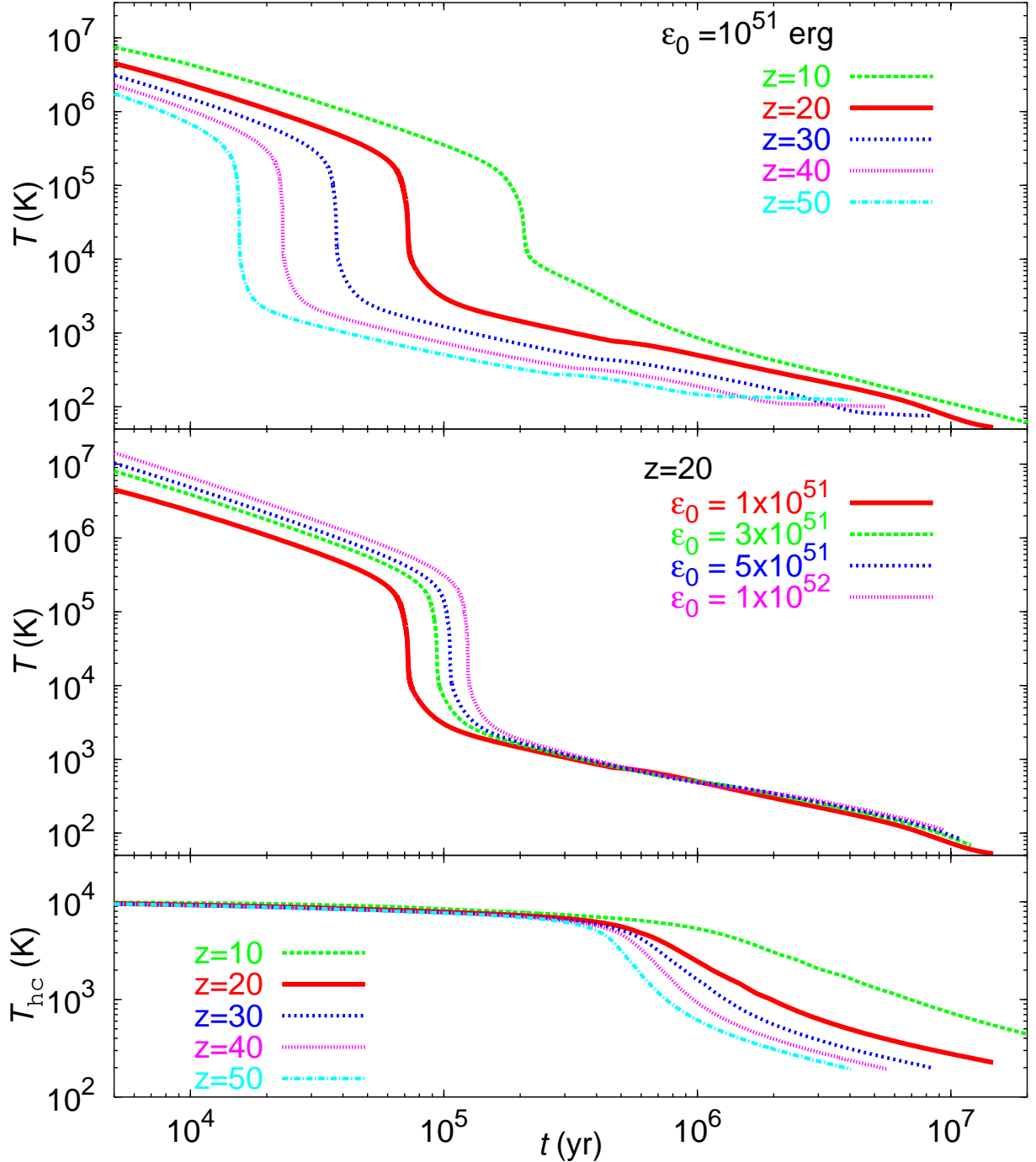}
\caption{
Time variations in the temperature of the gas shell (the top and middle panels) and that of the ambient gas (the bottom panel) are plotted against the elapsed time.
In the top and bottom panels, models with different ambient densities (or different redshifts of formation of host cloud $z = 10$, 20, 30, 40, and 50), but a fixed explosion energy ($\varepsilon_0 = 10^{51}$ erg), are compared.
The middle panel compares the models with different explosion energies ($\varepsilon_0$ = 1, 3, 5, and $10 \times 10^{51}$ erg), but a fixed ambient density (or redshift $z = 20$).
}
\label{fig:2}
\end{figure}

\begin{figure}
\plotone{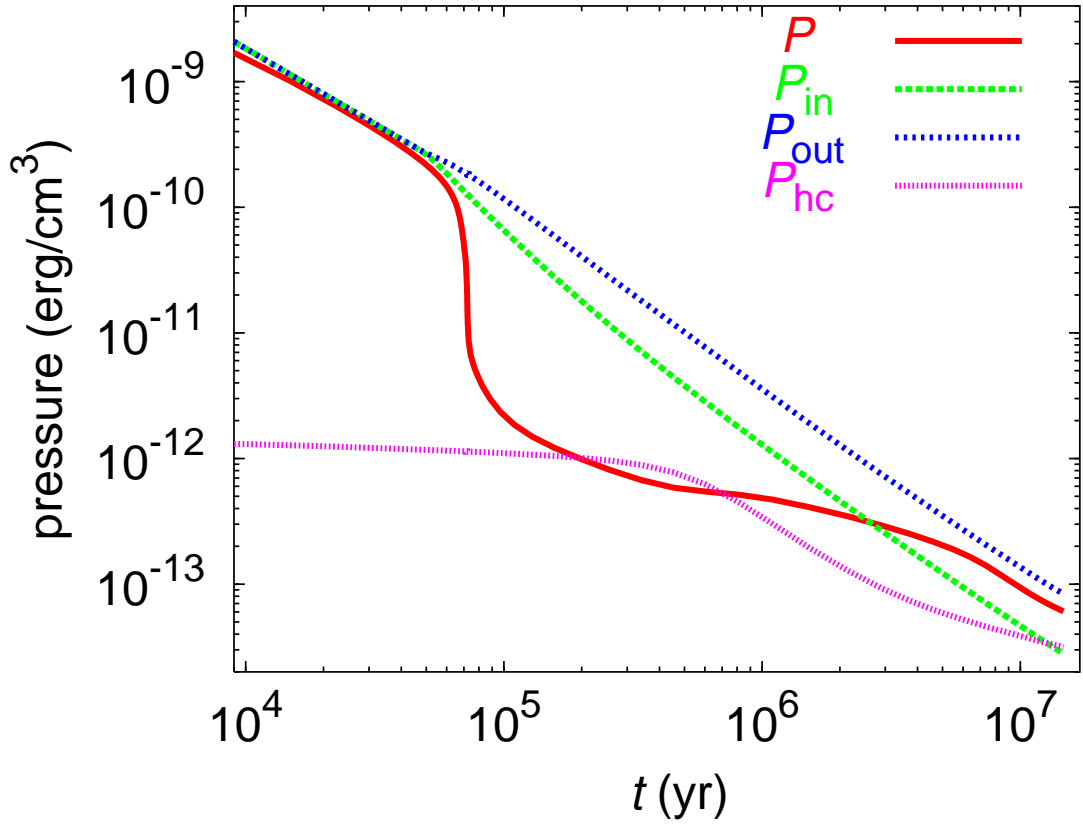}
\caption{
Variations of the pressures of the swept gas, $P$,
 the cavity gas, $P_{\rm in}$, the post-shock gas $P_{\rm out}$, and the ambient gas, $P_{\rm hc}$.
Model A20 of $(z, \varepsilon_0)=(20, 10^{51}$ erg) are shown for the ante-fragmentation phase.
}
\label{fig:3}
\end{figure}

\begin{figure}
\plotone{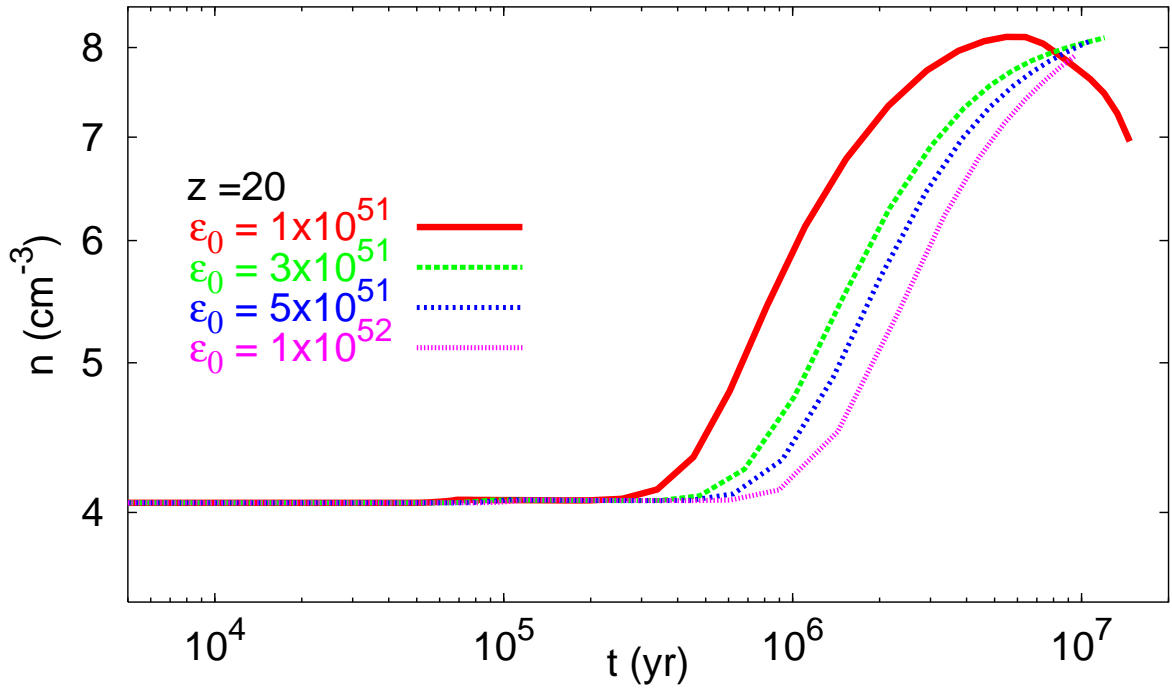}
\caption{
Time variations in the number density of gas in the gas shell for models with different energies ($\varepsilon_0= 1, 3, 5, \hbox{ and } 10 \times 10^{51}$ erg),  but a fixed redshift of $z = 20$. 
}
\label{fig:4}
\end{figure}

The evolutionary variations of thermal state in the swept gas shell and ambient gas are shown in Figures~\ref{fig:2}, \ref{fig:3}, and \ref{fig:4} as a function of elapsed time.
Figure~\ref{fig:2} shows the temperature variations of gas shell (top and middle panels) and ambient gas (bottom panel) for models with different redshifts (top and bottom panels) and for different explosion energies (middle panel).
The temperature of the SNR shell (top and middle panels) decreases gradually with  expansion during the Sedov-Taylor adiabatic stage.
The initial temperature is determined from the energy conservation at transition time $t_1$ in equation (\ref{eqn:initemp}).
The temperature steeply descends from $T\simeq 10^5$ K to $10^4$ K 
 owing to efficient atomic cooling by He and H (see Fig.~\ref{fig:7} left panel).
     As the temperature decreases below $10^4$ K, the temperature drop slows again, and yet, continues to decrease below 100 K owing to cooling by H$_2$ molecules and then by HD molecules.  
     There are two small dents discernible around $T \simeq 10^3$ K and 150 K, which are caused by the cooling due to H$_2$ and HD molecules, respectively.  
     For higher redshifts, the temperature decreases to the CMB temperature.  
     For $\varepsilon_0 = 10^{51}$ erg models, this occurs for $z \gtrsim 30$, and then, the SNRs evolve isothermally, as seen in the top panel of Fig.~\ref{fig:2} (at $t \gtrsim 10^6$ yr).
     For the larger explosion energy, however, the gas shell sweeps the
     ambient gas at a higher rate because of larger expansion velocity,
     and hence, can satisfy the condition for fragmentation at higher temperatures before it cools to the CMB temperature even for the largest redshift $z = 50$ as seen from the $ \varepsilon_0 \ge 1 \times 10^{52}$ erg models.   

Since the radiative cooling is in proportion to the density squared, the cooling is faster for  models with larger ambient densities (or larger redshifts).  
     Accordingly, the time interval from the supernova explosion to the shell fragmentation epoch is shorter for larger ambient gas density or for larger redshifts of the host cloud collapse, which decreases by a factor of $10 - 30$ from ${\rm several} \times 10^7$ yr to $\sim 10^6$ yr between $z = 10$ and 50.  
     If the ambient gas density is the same, the model with smaller explosion energy onsets the rapid atomic cooling earlier.  
     After the gas cools to $T \lesssim 10^3$K, the evolution converges because of the strong temperature dependence on atomic cooling.  
     Accordingly, the time interval from supernova explosion to fragmentation is elongated for a smaller explosion energy, since it takes longer to gather the necessary mass because of smaller expansion velocity.

In the bottom panel of Figure~\ref{fig:2}, we show the evolutions of the temperature, $T_{\rm hc}$, in the ambient gas for several different values of  collapse redshift $z$.
   The ambient gas evolves almost isothermally for $t \simeq 4 \times 10^5 \sim 10^6$ yrs, and then, the temperature decreases below $T_{\rm hc} = 1000$ K for $ 10^6 \sim 10^7$ yr. 
   For the model with $z=20$ and $\varepsilon_0 = 10^{51}$erg, the temperature of the ambient gas decreases to $T_{\rm hc} = 223$ K in 14.7 Myr by the time the temperature of the SNR shell decreases to $T = 53$ K and satisfies the conditions of fragmentation. 
   The cooling of the ambient gas is also faster for  models with higher redshifts because of higher density of ambient gas. 
   The ambient gas can finally cool to the temperatures $T_{\rm hc} = 193 - 656$ K when the condition of the fragmentation is satisfied, as listed in the 11-th column of Table~\ref{table:results} with the sound speed, $c_{\rm s, hc}$, in the 12-th column.

Figure~\ref{fig:3} illustrates the time variations in the pressure in the gas shell, $P$, along with those of the boundary pressures, i.e., the cavity pressure inside the gas shell, $P_{\rm in}$ (eq.~[\ref{eq:pin}]) and the post-shock pressure outside the gas shell $P_{\rm out}$ (eq.~[\ref{eq:pout}]) for the model with $z=20$ and $\varepsilon_0 = 10^{51}$ erg. 
     As the atomic cooling becomes very effective in the pressure-driven stage, the pressure in the shell starts to decrease nearly in proportion to the temperature in the gas shell since $\tau_{\rm cool} \ll \tau_{\rm dyn}$ and hence, the dynamical readjustment of structure cannot catch up with the temperature drop caused by the radiative cooling.   
     The radiative cooling rate declines very steeply below $10^4$ K.  
     As the cooling timescale grows longer and eventually exceeds the dynamical timescale ($\tau_{\rm dyn} \ll \tau_{\rm cool}$), the gas shell starts to readjust its structure to the surrounding pressures, and the pressure approaches the average of $P_{\rm in}$ and $P_{\rm out}$ according to equation (\ref{eq:dyncool}).  
     In particular, for $t \gtrsim 10^6$ yr, when the time grows longer than $\tau_{\rm dyn}$, the gas shell restores the adjustment and the pressure $P$ tends to be controlled by the boundary pressures and decreases with them.

In this figure, we also plot the ambient pressure $P_{\rm hc}$. 
   The ambient pressure remains much lower than the inner pressure, $P_{\rm in}$, for most of the time, and it is only after $t > 1.3  \times 10^7$yr that the ambient pressure grows higher than the inner pressure.  
   Since their difference attains only 9 \% at the time of fragmentation and is much smaller than the momentum flux of the gas shell, the effect of decelerating the expansion of the SNR shells is rather small, and hence, will not affect the evolution of the SNR shell so much. 
   In this model, the SNR gas shell keeps expanding at a velocity exceeding the sound speed, and hence, undergoes fragmentation without dissolving and merging into the ambient gas. 
   For the models of larger explosion energy and higher ambient density, the condition that $V  > c_{\rm s, hc}$ holds by a greater margin because of greater expansion velocity at the time of fragmentation, as seen from Table~\ref{table:results}.

The changes in the number density in the gas shell are plotted in Figure~\ref{fig:4}.   
     The gas density in the shell is kept nearly constant at $4 \rho_0$ until $\tau_{\rm dyn}$ becomes shorter than $\tau_{\rm cool}$ at $t \gtrsim 3 \times 10^5$ yr and for $T \lesssim 10^4$ K.  
     Then, the density starts to increase as the gas shell undergoes compression by the pressures at the inner and outer boundaries.  
     Finally, when the gas shell restores the hydrostatic equilibrium, the density begins to decrease along with the boundary pressures that confine the gas shell, as seen from the model with the smallest explosion energy.  
     For the greater explosion energy, however, this final stage does not occur;  
     because of greater expansion velocity, the SNR gathers more gas to make the shell thicker and the dynamical timescale longer in proportion, and hence, the fragmentation condition is satisfied before the hydrostatic equilibrium is restored.    

\begin{figure}
\plotone{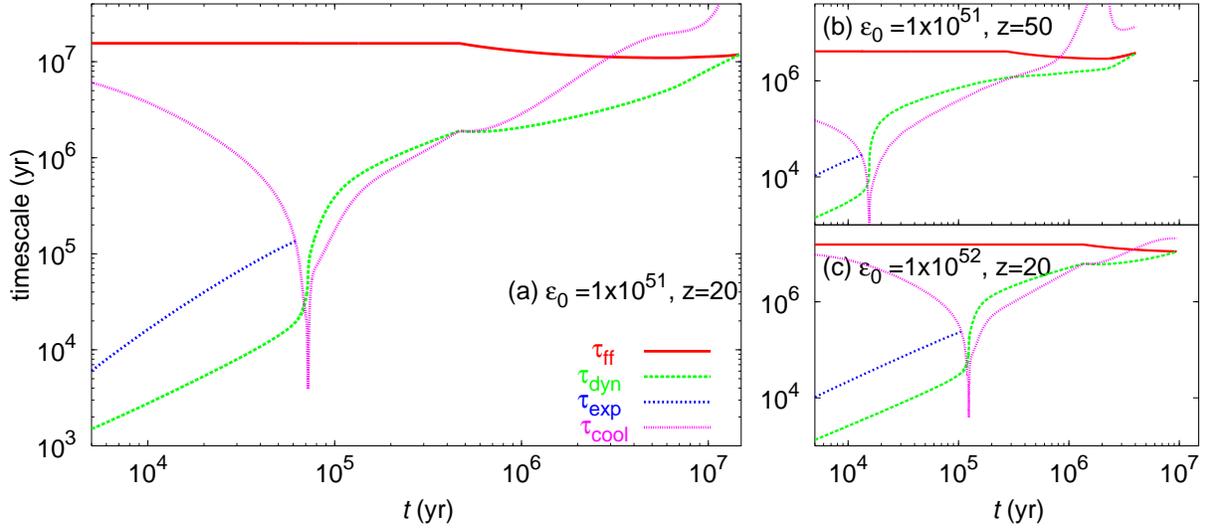}
\caption{
Evolutions of four typical timescales ($\tau_{\rm ff}$: free-fall timescale, $\tau_{\rm dyn}$: dynamical timescale, $\tau_{\rm exp}$: expansion timescale, and $\tau_{\rm cool}$: cooling time scale) are plotted against the elapsed time.
The left panel is for a model with ($z, \varepsilon_0$)=(20, 1$\times$10$^{51}$ erg), and top-right panel is for a model with ($z, \varepsilon_0$)=(50, 1$\times$10$^{51}$ erg) and the bottom-right panel is for a model with ($z, \varepsilon_0$)=(20, 5$\times$10$^{51}$ erg).
$\tau_{\rm exp}$ is shown only for the Sedov-Taylor stage.
}
\label{fig:5}
\end{figure}

We exemplify the behaviors of the four timescales, mentioned above, in Figure~\ref{fig:5}, which shows the evolutionary changes in the expansion time, the cooling time, the dynamical time, and the free-fall time, mentioned above for three models of ($z$, $\varepsilon_0$) = (20, $10^{51}$ erg; left panel), (50, $10^{51}$ erg; top-right panel) and  (20, $10^{52}$ erg; bottom-right panel).  
     As the temperature descends to $T \simeq 10^5$ K, $\tau_{\rm cool}$ starts to decline rapidly owing to  atomic cooling, and eventually becomes shorter than the expansion timescale $\tau_{\rm{exp}}$ which increases with time.  
     This causes a transition from the Sedov-Taylor adiabatic stage to the pressure-driven stage.  
     When the temperature falls further  to $T \simeq 2 \times 10^4$ K, $\tau_{\rm cool}$ reverses the direction of the change and begins to increase.  
     As the recombination of hydrogen atoms proceeds with a further drop in the temperature, $\tau_{\rm cool}$ grows large very rapidly, increasing by a factor of $\sim 100$ while the temperature decreases to $T \simeq 5 \times 10^3$ K, and forms a sharp minimum.   
     Thereafter, the cooling timescale continues to increase constantly, slowing the increase rate slightly by the enhancement of cooling by H$_2$ and HD molecules (two small dents are discernible at $t \simeq 5 \times 10^5$yr and $5 \times 10^6$ yr in the left panel).
     On the other hand, the dynamical timescale grows large owing both to the decrease in the temperature and to the increase in the shell thickness.  
     When the atomic cooling begins to be effective, therefore, the cooling timescale becomes smaller than the dynamical timescale.  
     As the temperature decreases below $10^4$ K, they reverse their relation and the cooling timescale grows larger than the dynamical timescale again.  
     When $\tau_{\rm dyn} \lesssim \tau_{\rm cool}$, the gas shell undergoes compression, as seen above. 
     Then the density in the shell increases, which causes a decrease in the free-fall timescale, as seen from $t \gtrsim 5 \times 10^5$ yr, while the dynamical timescale continues to increase because of the decrease in the shell temperature, and also because of the increase in the shell thickness.      
     Finally, $\tau_{\rm ff}$ becomes as short as $\tau_{\rm dyn}$ ($t \simeq 10^7$ yr for the left panel), and the gas shell is expected to break up into fragments, each of which undergoes  gravitational contraction. 

For higher ambient density (or higher $z$; top-right panel), the cooling, dynamical, and free-fall timescales are shorter, while the expansion timescale is 
 slightly dependent on the density in the gas shell.
These accelerate the evolution.  
     Accordingly, the fragmentation epoch becomes earlier, because the gas in the shell cools more rapidly and the free-fall timescale is smaller for higher density. 
For the model with greater explosion energy (bottom-right panel),
 the temperature remains higher, which delays  transition
 to the pressure-driven stage. 
In the later stage, the difference in the temperature tends to be eliminated as a result of the atomic cooling.  
     On the other hand, faster expansion makes the shell thicker and the dynamical timescale longer.  
     This delays the arrival of the epoch of $\tau_{\rm cool} > \tau_{\rm dyn}$ and the contraction of the gas shell, as seen in Figure~\ref{fig:4}, but brings forward the epoch of fragmentation slightly.  
     Accordingly, we conclude that the time necessary for fragmentation depends significantly on the ambient gas density ($\rho_0$ or the formation redshift $z$), but weakly on the explosion energy ($\varepsilon_0$).

\begin{figure}
\plotone{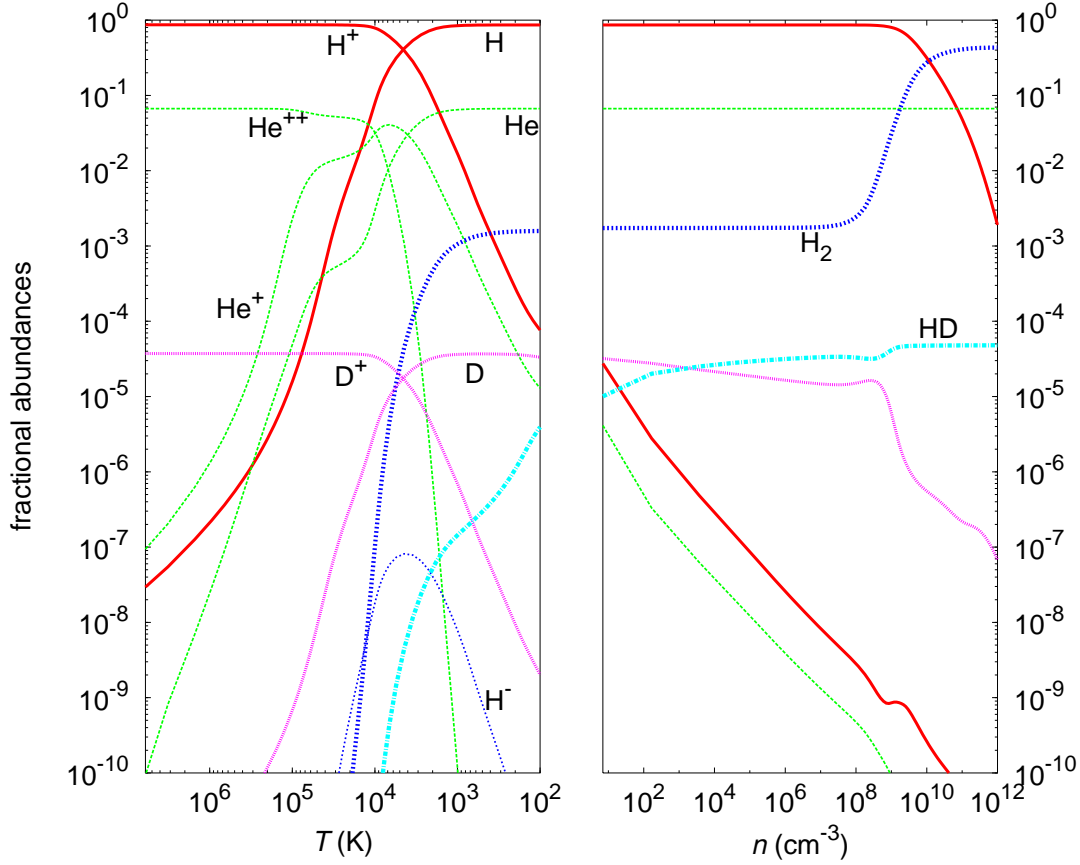}
\caption{
Time variations in the fractional abundances for 10 species (H, H$^+$, H$^-$, H$_2$, He, He$^+$, He$^{++}$,  HD, D, and D$^+$) in the gas shell for the model with $\varepsilon_0 = 10^{51}$ erg and $z = 20$.
   The abundance is plotted as a function of temperature (left panel) before the fragmentation, and as a function of number density (right panel) after the fragmentation.
In the post-fragmentation phase,
 we assume a cylindrical fragment contracts in the radial direction.
}
\label{fig:6}
\end{figure}

\begin{figure}
\plotone{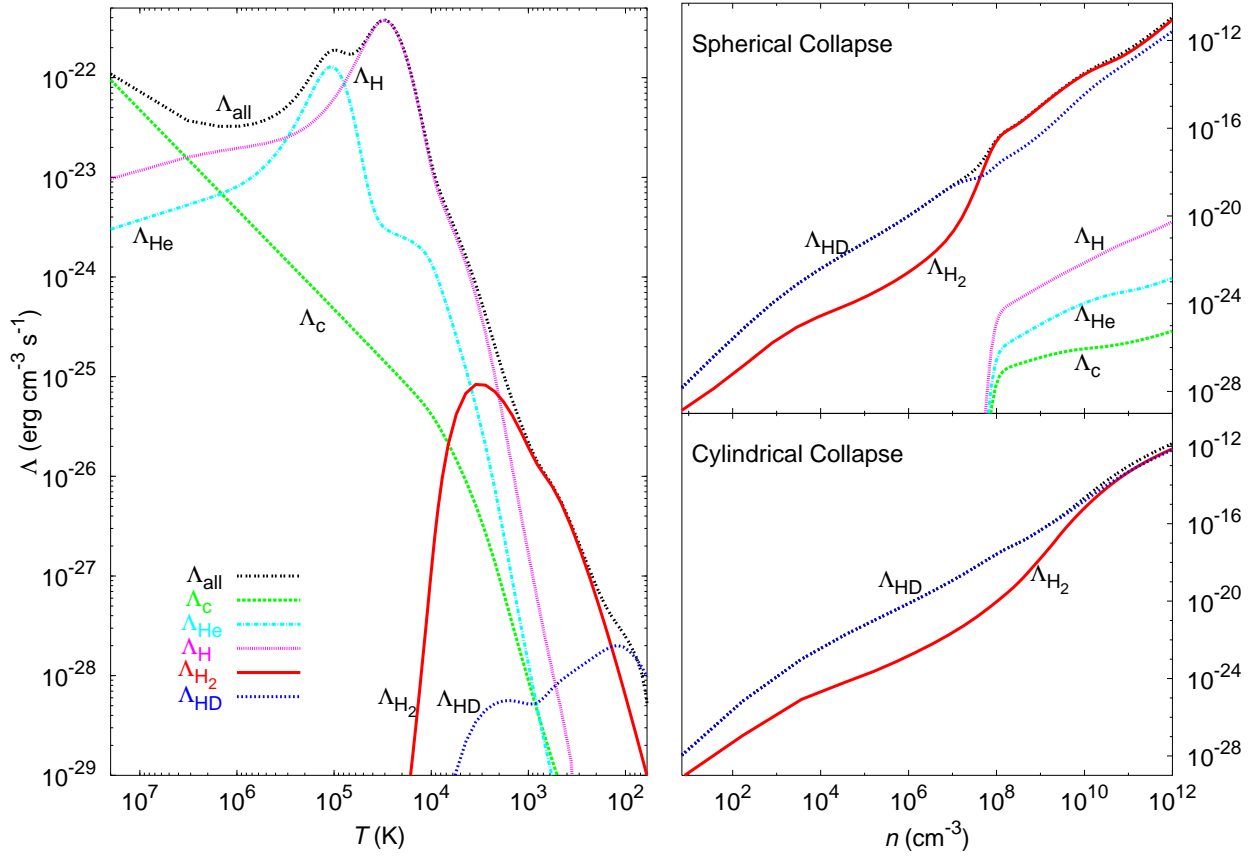}
\caption{
Cooling rates accomplished in the gas shell ($\Lambda_{\rm all}$: total cooling rate, $\Lambda_{\rm c}$: inverse Compton cooling, $\Lambda_{\rm H}$: Hydrogen cooling, $\Lambda_{\rm He}$: Helium cooling, $\Lambda_{\rm H_2}$: molecular Hydrogen cooling, and $\Lambda_{\rm HD}$: HD molecular cooling) are plotted against the gas temperature in the ante-fragmentation phase (left panel) and against the number density in the post-fragmentation phase (right two panels).
The top-right panel indicates the cooling rates in the case of spherical collapse, and the bottom right is for the case of cylindrical collapse.
This corresponds to the model with $(z,\varepsilon_0)=(20, 10^{51} {\rm erg})$. 
}
\label{fig:7}
\end{figure}

We present  typical patterns of the evolution of fractional abundances and of the variations in the cooling rates against the shell temperature, $T$, in the left panel of Figures~\ref{fig:6} and \ref{fig:7}, respectively, for the $(z, \varepsilon_0) = (20, 10^{51}\hbox{ erg})$ model.  
     The ambient gas is ionized by the SNR shock at first, and then, recombines around $T \simeq 5 \times 10^3$ K. 
     Thereafter, H$_2$ and HD molecules are formed through the reactions as
\begin{eqnarray}
\rm{e}^- + \rm{H} \rightarrow \rm{H}^- + h\nu,   \\
\rm{H}^- + \rm{H} \rightarrow \rm{H}_2 +e^-,
\end{eqnarray} 
and 
\begin{equation}
\rm{D}^+ + \rm{H}_2 \rightarrow \rm{HD} + \rm{H}^+.
\label{eq:HDmake}
\end{equation}
Since H$^-$ ions react as a catalyst to make hydrogen molecules, both H$_2$ and HD abundances increase while H$^-$ abundance increases, as in the left panel of Figure~\ref{fig:6}.  
The HD abundance continues to increase even after the H$^{-}$ abundance decreases and the H$_2$ abundance saturates. 
This is due to the charge transfer reaction in equation (\ref{eq:HDmake}), through which HD molecules are formed from much more abundant H$_2$ molecules.  

The left panel of Figure~\ref{fig:7} plots the contributions from the cooling rates $\Lambda \ (T, n, {\rm Composition})$ for the same model as in Figure~\ref{fig:6}.  
This elucidates the fact that H$_2$ and HD molecules are effective coolants below the temperature $\sim 2 \times 10^3$ K and $\sim 150$ K, respectively. 
The cooling rate is smaller by a factor of $10^3$ and $10^6$ than the atomic cooling rate at its peak.  
If it were not for H$_2$ and HD molecules, however, gas could not be cooled below $\lesssim 10^4$ K within the Hubble time. 
In the model presented in Figure \ref{fig:7}, the fractions of H$_2$ and HD molecules amount to $n({\rm H}_2)/n_{\rm 0,H} = 1.58 \times 10^{-3}$ and $n({\rm HD})/n_{\rm 0,H} = 8.32 \times 10^{-6}$ at the fragmentation epoch, which are much larger than their primordial values of $10^{-6}$ and $10^{-9}$, respectively \citep{gal98}. 
These abundances agree well with the result of Uehara \& Inutsuka (2000).  
When the temperature becomes as low as 145 K, the HD cooling exceeds the H$_2$ cooling and promotes a further drop in gas temperature.  
The fragmentation condition, $t_{\rm ff} = t_{\rm dyn}$, is satisfied only when the temperature descends to $T = 53$ K at the age of 14.7 Myr after the SN explosion.   
By this stage, the shell expands to a radius of 90 pc and the gas shell has gathered the ambient gas of mass of $1.11 \times 10^5 \msun$ for this particular model. 

\subsection{The Post-Fragmentation Phase}
After the condition of $\tau_{\rm ff} = \tau_{\rm dyn}$ is fulfilled, we assume fragmentation to occur.  
Then the evolution proceeds to the final stage of fragmentation and star formation. 
We follow the gravitational contraction of the fragment pieces assuming two different types of fragment geometry: sphere and cylinder.

\begin{figure}
\plotone{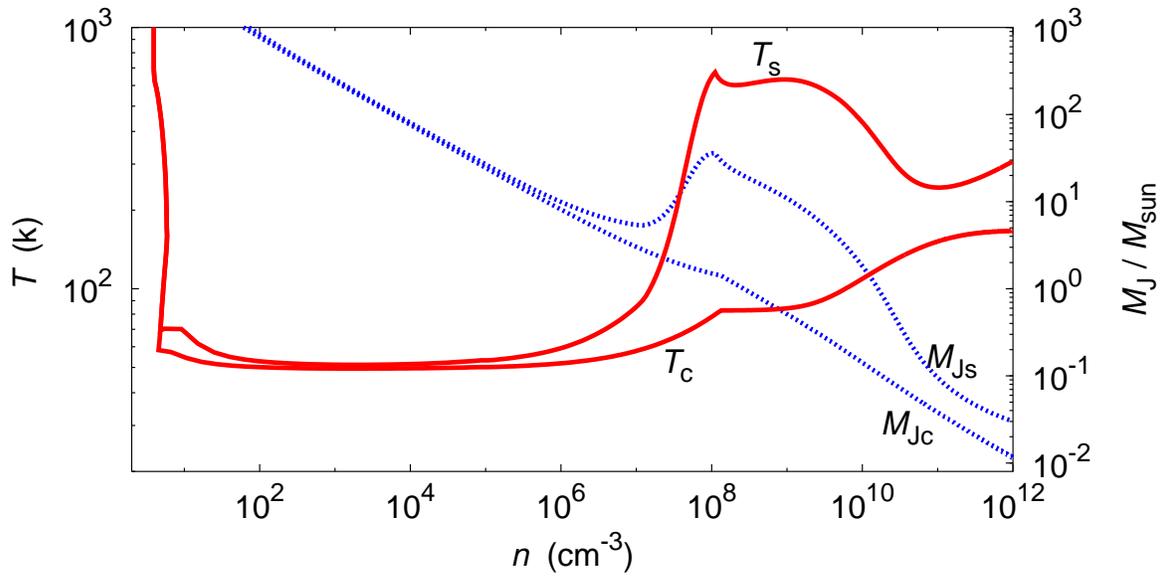}
\caption{
Evolution of gas temperature ($T$: solid curves)
 and the Jeans mass ($M_J$: broken curves) in the fragments during
 the post-fragmentation phase for a model with  ($z$, $\varepsilon_0) = (20, 10^{51}$ erg). 
Two geometrical models of collapse are shown for spherical ($T_s$, $M_{\rm Js}$)
 and cylindrical ($T_c$, $M_{\rm Jc}$) collapses.
}
\label{fig:8}
\end{figure}

The temperature evolution during the post-fragmentation phase of the$(z, \varepsilon_0) = (20, 10^{51}$ erg) model is shown in Figure~\ref{fig:8} for both the spherical and cylindrical collapses.  
For the cylindrical case, the gas temperature $T_c$ decreases nearly to, but slightly higher than, the CMB temperature.
Fragmentation occurs at the age of 14.7 Myr after the SN explosion.
At this epoch, $T_c\simeq 53$K and $n_c \simeq 7{\rm cm}^{-3}$, which corresponds to a steep break near the lower-left corner in Figure~\ref{fig:8}.
Before the fragmentation, the evolutionary path in Figure \ref{fig:8} is vertically downward, that is, gas cools without strong compression.
After this epoch, the fragment changes its evolution as it evolves nearly isothermally. 
The temperature of the fragment decreases further to 48K, which decreases  the critical line-mass $\lambda_c\propto T_c$ by 19\% in 6.7 Myr from the fragmentation epoch.
The shell expands from $R \simeq 90$ pc at $t=14.7$ Myr to $R \simeq 103$ pc at  $t=28.7$ Myr, which increases the line-mass $\lambda$ approximately 50\%.
These two factors work cooperatively to promote further collapse of the fragment. 
Thereafter, the fragment evolves under the condition that $\lambda_{\rm c} \simeq \lambda$.

The gas temperature remains nearly constant for the number density
 $n \lesssim 10^6 \hbox{ cm}^{-3}$, and then,
 increases  to $T_c \sim 100$ K
 owing to the increase in the mean molecular weight,
 as also seen from \citet{uehara00}.  
     Variations in the fractional abundances are shown in the right panel of Figure~\ref{fig:6} against the number density in the fragment pieces for the cylindrical contraction case.  
     As the fragment contracts, the HD fraction gradually increases with density through the reaction in equation~(\ref{eq:HDmake}).  
     The H$_2$ fraction changes little until the density becomes sufficiently high for the three-body reactions ($n \gtrsim 10^8 \rm{cm^{-3}}$).
A break near $n\simeq 10^8{\rm cm}^{-3}$ in Figure \ref{fig:8}
 corresponds to this critical density of H$_2$.
Beyond that density, the H$_2$ fraction starts to increase rapidly with density. 
     The variations in the cooling rates are also plotted in the right panels of Figure~\ref{fig:7} against the number density for both the spherical and cylindrical cases.  
     H$_2$ and HD molecules are only the coolants effective after the fragmentation.
It is clearly shown that HD molecules play a more important role in the lower temperature.  
    HD molecules continue to dominate the cooling throughout the post-fragmentation phase, which allows the contraction of the fragments at much lower temperatures than found in the previous computations involving only H$_2$ molecules \citep{puy93,uehara96,nakamura99,flower03}.  

For the spherical case, on the other hand, the temperature keeps rising with density owing to dynamical compression (Fig.\ref{fig:8}).  
     When the temperature becomes higher than $T_s\simeq$ 150K, HD molecules are dissociated to decrease the fractional abundance by a factor of $\sim 1/10$, and hence, the cooling rate due to HD molecules is overtaken by that due to H$_2$ molecules for $n \geq 10^8 \hbox{ cm}^{-3}$.  
     Thereafter, the H$_2$ fraction increases through the three-body reactions, which enhances the contribution of H$_2$ molecules to the cooling.  
     Although the HD fraction also increases again through the charge exchange reaction with the H$_2$ molecules, the latter continues to dominate the cooling because of high temperatures.  
    The temperature in the fragment pieces is, however, kept much lower than obtained in the previous computations without the HD molecules because of the lower temperatures at the fragmentation epoch \citep{ferrara98}.  

We also show the time variations in the Jeans masses in  fragment pieces during the post-fragmentation phase in Figure~\ref{fig:8}.  
     For the cylindrical case, the Jeans mass $M_{\rm Jc}$ decreases to $0.16 \msun$ at the stage when fragment pieces become optically thick at the density $n \simeq 10^{10} \rm{cm}^{-3}$.  
     After this stage, the cooling becomes ineffective and the gas cannot contract in the mass-scale smaller than the Jeans mass, while the gas element more massive than the Jeans mass may collapse in a free-fall timescale.  
     Because an adiabatic core is formed,  further fragmentation can give rise to stars with the Jeans mass at the core formation epoch, which is well below one solar mass and corresponds to a less massive star which can survive to date.
We note that our Jeans mass is larger than that of Uehara \& Inutsuka (2000),
$M_{\rm Jc} \sim 0.04\,\msun$ , because they seem to underestimate the optical 
depths of HD lines (see e.g., Nakamura \& Umemura 2002).  
Therefore, our fragments are likely to evolve into low-mass stars, rather than primordial brown dwarfs.
     For the spherical case, on the contrary, the fragment may not stop contracting even after it becomes optically thick \citep[e.g. ][]{omukai98}.  
However, since the gas shell is thin
 (the ratio of the thickness to the radius ${}\simeq 0.1$),
 the fragment pieces are expected to form a cylindrical
 filament whose axis is parallel to the shell,
 and hence, can give rise to stars with sub-solar masses.  

\section{Fragmentation Conditions in the Primordial Clouds}
In the preceding section, we have solved the evolution of  gas shells swept by the supernova remnants of  first-generation stars, assuming that the host clouds have sufficiently large masses. 
In order for  star formation to be actually triggered in host clouds, the following two conditions have to be satisfied at the fragmentation epoch:
 i.e., (1) the shell expansion velocity is larger than the sound speed of ambient gas and (2) the mass of swept gas is smaller than the total baryon mass of the host cloud.
In our calculations, we have shown that the SNRs can give rise to  low-mass star formation while the expansion velocity is larger than the ambient sound speed, and hence, without dissolving into the interstellar gas, except for the two cases of weakest explosion energy of $\varepsilon_0 = 1$ and $3 \times 10^{51}$ erg at the lowest collapse redshift of $z=10$ (see $V$ and $c_{\rm s, hc}$ in the 7th and 12th columns in Table~\ref{table:results}).  
     Then, the fragmentation occurs in the host cloud, if the baryon mass of the host cloud is larger than the mass, $M_{\rm sw}$, of the gas swept by a SNR at the fragmentation epoch.   
     If the expansion velocity is slowed  to the ambient sound speed, the gas shell dissolves into the ambient gas. 
     If a sufficient mass of gas is not available, SNRs expand beyond the edges of host clouds, either falling back and being mixed into the interstellar clouds or being dispersed out of clouds and spread into the intercloud space, depending on whether the expansion velocity is lower or greater than the escape velocity of the host clouds.  
     We discuss below first the dependence on the mass of host clouds and then the effect of the ambient temperature.  

The values of swept gas mass, $M_{\rm sw}$, at the fragmentation epoch are given in Table~\ref{table:results}, which gives the minimum baryon masses of the host cloud necessary for the fragmentation under a given set of the ambient gas density (or collapse redshift) and the explosion energy.   
This mass varies by a factor of $\sim 10$ in the parameter range we computed from $6.07 \times 10^4$ to $8.04 \times 10^5 \msun$, and increases with the explosion energy and decreases for higher ambient gas density (or larger redshift at the collapse of host cloud).
This means that less massive host clouds can form low-mass stars only when they are formed in the earlier epoch (higher $z$ and thus larger $\rho_0$).
 
\begin{figure}
\plotone{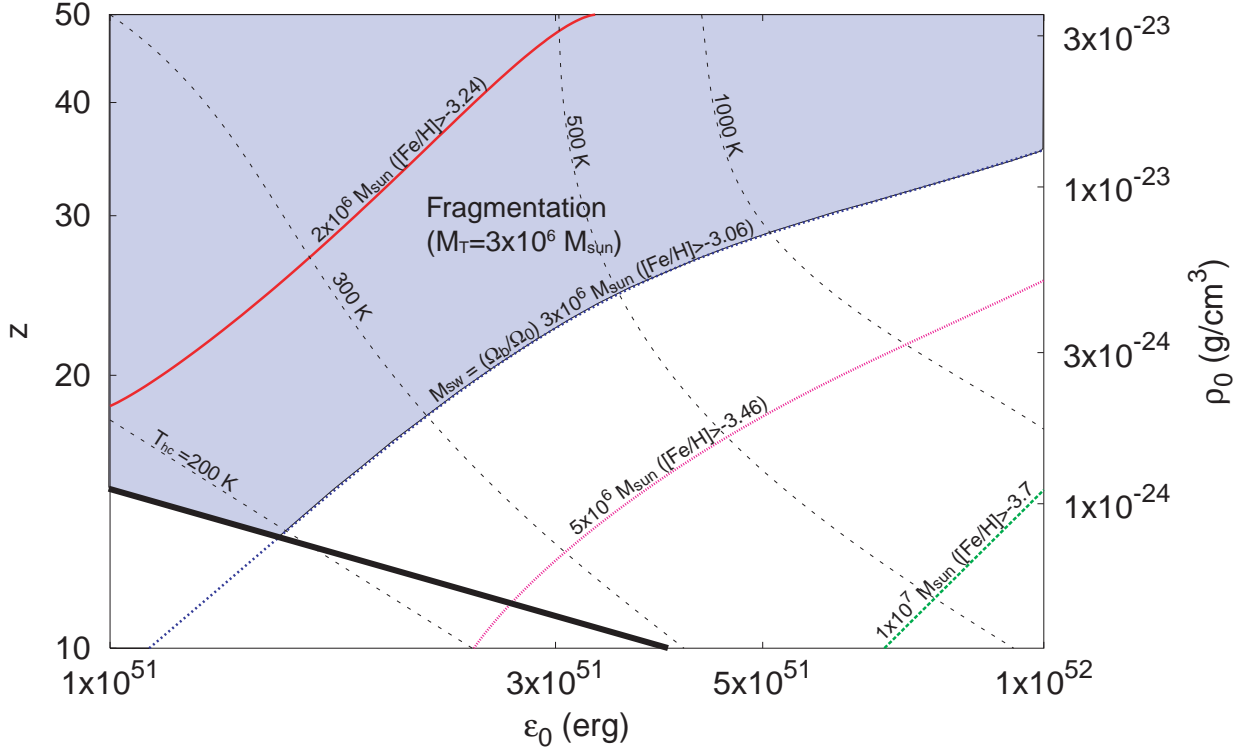}
\caption{
Lower bounds of the redshift $z$ (the left axis) or of the background density, $\rho_0$, in the host clouds (the right axis) for star formation to be triggered by a single supernova explosion are plotted as a function of the explosion energy, $\varepsilon_0$ of supernova for the host clouds of total mass, given mass $M_{\rm T}$= 2, 3, 5, and  $10 \times 10^6 \msun$, from the top to the bottom. 
   The thick line denotes the border of $V = c_{\rm s, hc} (T)$,  which divides whether the gas shell swept by SNR can fragment or merge into the ambient gas, with the temperature in the ambient gas derived by solving the chemical reaction and thermal evolution of the ambient gas simultaneously. 
E.g., the shaded region denotes the parameter range in which we expect the SNR-induced fragmentation of low-mass second-generation stars in the host clouds with the total mass of $M_{\rm T}=3\times 10^6M_\odot$. 
   Dotted lines represent the dependence on the shell expansion speed (or the temperature)in the ambient gas for the temperatures of 200 K, 300 K, 500 K, and 1000 K. 
They give the boundaries for the gas shell, swept by SNR, to fragment without mixing into the ambient matter, i.e., supernova-triggered, low-mass star formation is expected in the parameter range to the right of these lines, and requires a supernova of larger explosion energy for higher temperatures in the ambient gas.  
   For the supernova of weaker energy, the SNR shell dissolves into the ambient gas before the fragmentation. 
}
\label{fig:9}
\end{figure}

The necessary baryon mass, $M_{\rm sw}$, of the host cloud for the fragmentation can be converted into the total (baryon plus dark matter) mass of the host clouds by assuming an appropriate model of the Universe. 
In Figure~\ref{fig:9}, we illustrate the permitted region (i.e., $\hbox{the baryon mass} \ge M_{\rm sw}$) in the $\varepsilon_0-z$ plane for various total masses by assuming the Einstein-de-Sitter model, i.e., $\Omega_0 = 1$, $\Omega_{\rm B} = 0.06$, and $M_{\rm T}=M_{\rm sw}(\Omega_0/\Omega_b)$.  
For a given mass of the host cloud, there exists a lower limit to the ambient gas density or the redshift of the supernova explosion that allows fragmentation of the gas shell;  the lower limit is an increase function of $\varepsilon_0$.
   In the case of the first collapsed objects with total masses of $\sim 3\times10^6 \msun$ (see the hatched region of Fig.\ref{fig:9}), fragmentation can occur within the host cloud in the hatched area of narrow parameters, e.g., if the supernova of $\varepsilon_0$ = 3$\times 10^{51}$ erg exploded at $z \gtrsim 20$ or that of $\varepsilon_0 = 10^{52}$ erg exploded at $z \gtrsim 30$.  
Because of the rather narrow range of $M_{\rm sw}$, mentioned above, the mass of primordial clouds that can sustain  supernova-triggered star formation is bounded sharply between the total mass of cloud $M_{\rm T} \simeq 10^6 \sim 10^7 \msun$. 
That is, for the host clouds of $M_{\rm T} = 2 \times 10^6 M_\odot$,  supernova-triggered star formation is possible only in a narrow parameter region of $z \ge 20$ and $\varepsilon_0 < 3 \times 10^{51}$ erg. 
For the host clouds of $M_{\rm T} > 10^7 M_\odot$,  star-formation can always be triggered, except for the narrow range in the right-bottom corner of $z \simeq 10$ and $\varepsilon_0 \simeq 10^{52}$ erg.    

We have also solved the thermal evolution of ambient gas consistently with the effect of the first generation stars taken into account, to show that the temperature in the ambient gas decreases efficiently to the range of $T_{\rm hc} = 193 - 656$ K by the time the fragment condition is satisfied (Table~\ref{table:results}). 
   By use of these temperatures, we may derive the condition that  $V  > c_{\rm s, hc} (T_{\rm hc})$, which is plotted by a thick line in the left bottom corner in Fig.~9. 
   In the parameter region above and to the right of this line, i.e., if a supernova explodes with a larger explosion in the host cloud of higher density than delineated by this line, the gas shell swept by the SNR can fragment to give birth to low-mass stars surviving to date. 
   Otherwise, the swept gas shell will dissolve and merge into the ambient gas without triggering fragmentation.

In the above computations, we take into account the heating of ambient gas only by the SN progenitor, which ends with the supernova explosion. 
   The temperatures in the ambient gas may vary and can be higher, however, if gas is irradiated by other heating sources, such as other first-generation massive stars or the SNR itself, and/or if the SNR shell expands beyond the Str\"omgren sphere, in which the temperature has not been raised high enough to ionize gas, and hence, the subsequent cooling is not so effective. 
  In order to see the dependence of the evolution and fragmentation of the SNR gas shell on the thermal state of ambient gas, we calculate the evolution of several SNR models with the temperature of ambient gas set at a constant value, i.e., $T_{\rm hc} = T_0$. 
   For $z=20$ and $\varepsilon = 10^{51}$, the gas shell expands and cools enough to meet the fragment condition $\tau_{\rm ff} = \tau_{\rm dyn}$ when the SNR radius reaches 92 and 85 pc in the models of constant ambient temperature $T_{\rm hc} = 200$ and $10^3$ K, respectively.  
   The difference in the radii reflects the effect of deceleration by the ambient gas pressure.  The condition under which  the swept gas shell keeps expanding without being dissolved into the ambient medium ($V > c_{\rm s, hc}$) is satisfied in the model of the lower ambient temperature, consistent with the above result of the calculation which solves the thermal evolution of ambient gas simultaneously. 
   In the model of $T_{\rm hc}= 10^3$ K, this condition is violated earlier, and hence, the gas shell has to be disturbed and merges into the ambient medium before it can gather gas sufficient to trigger the fragmentation. 
   For such a high ambient temperature as $T_{\rm hc} = 10^4$ K, the expansion of the SNR is halted at the radius of 52 pc because of the high pressure of the ambient medium, and the swept gas shell will be dissolved and melted in the host cloud.  

In Figure~\ref{fig:9}, we also show the parameter region, where the conditions of $V  = c_{\rm s, hc} (T_{\rm hc})$ are satisfied until the fragmentation epoch, by dotted lines for the different ambient temperatures of $T_{\rm hc} = 200$, 300, 500 and $10^3$ K. 

In the parameter range left of these lines, where $V  < c_{\rm s} (T_{\rm hc})$, the SNR shell is dissolved into the ambient medium before the fragmentation epoch for a given ambient temperature $T_{\rm hc} = T_0$. 
   If the ambient temperature is kept higher, the fragmentation demands a supernova of greater explosion energy, since the SNR has to sweep the amount of ambient gas necessary for  fragmentation before the expansion is decelerated to  sound velocity in the ambient gas.  
For example, if the ambient medium has been cooled to 300 K by radiative cooling, the explosion energy of $\varepsilon_0 \gtrsim 2 \times 10^{51}{\rm erg}$ is necessary for $V  > c_{\rm s, hc}(300{\rm K})$ in a host cloud of $M_{\rm T} = 3 \times 10^6 \msun$ at $z=20$, while for the cloud with  $T_{\rm hc}=1000 {\rm K}$, a much stronger explosion energy of $\varepsilon_0 \gtrsim 5 \times 10^{51}{\rm erg}$ is necessary for the SNR shell to fragment. 
   For a high temperature of $T_{\rm hc} \sim 10^4{\rm K}$, such as assumed by \citet{sal03},  SN-triggered star formation is possible only for a supernova (or hypernova) with an explosion energy exceeding $10^{52}$ erg, consistent with their results. 
   But this is not the case, since it is likely that gas in the host cloud has been cooled after the supply of UV photons from the SN progenitor.

\section{Conclusions and Discussion}
In this paper, we have addressed two questions whether a primordial SNR can trigger star formation in a first collapsed object and whether low-mass stars can be formed to survive to date.  
For that purpose, we have studied the evolution of supernova remnants in primordial clouds with H$_2$ and HD chemistry taken into account for a range of parameters of ambient density (or the redshift) of host clouds and  explosion energy of supernovae, as summarized in Table \ref{table:results}.  
It is found that a gas shell swept by a supernova shock undergoes cooling and that HD molecules can effectively cool the temperature to as low as $32 - 154$ K.
Fragmentation begins when the mass in the gas shell reaches the range of $6.1 \times 10^4 \msun - 8.0 \times 10^5 \msun$.  
This is a necessary mass, $M_{\rm sw}$, for star formation to be triggered by SNRs in  primordial clouds of total mass $M_{\rm T}$.
$M_{\rm sw}$ depends on the ambient density (or the redshift) and the SN explosion energy.   
The standard CDM model predicts that the masses of the collapsed objects from the 3 $\sigma$ peaks vary from $M_{\rm T} \simeq 2 \times 10^6 \msun$ at $z = 30$ to $\simeq 10^8 \msun$ at $z = 10$  (e.g., \citealt{haiman96}, \citealt{tegmark97}), which has a baryon mass of
 $M_b\simeq 10^5\msun$ ($z=30$) and $M_b\simeq 5\times 10^6\msun$ ($z=10$).  
     If $M_{\rm sw} < M_b = \Omega_{\rm B} M_{\rm T}$, SNR can fragment within the host cloud to bring about  supernova-triggered star formation.   
     Otherwise, the gas shell passes  the edge of the host cloud, and hence, is blown out without  fragmentation.  

In the range of parameters studied, fragmentation occurs in the
     pressure-driven expansion stage.
     The SNR expands to radii of $30 \sim 332 $ pc by this stage, which increases with the explosion energy and decreases with the ambient density.  
     It takes $1.1 {\rm Myr} - 45 $ Myr from the supernova explosion, much shorter than the Hubble age at that epoch, for the swept gas shell to start fragmentation, which is shorter for higher ambient density and for larger supernovae explosion energy.    
The lifetimes of massive stars of
 $M \simeq \hbox{several tens} \ M_\odot$ as $\tau\simeq 10{\rm Myr}$ 
 are longer than the elapsed time for high $z$ SNR ($t_f\lesssim 5{\rm Myr}$ for $z=50$ models) and at least comparable for the supernova explosion at lower redshift
($t_f\simeq 10{\rm Myr}$ for $z=20$ models).   
     Accordingly, it may be possible that fragmentation occurs even before other subsequent supernovae in the same primordial clouds  contaminate the swept gas shell.

\subsection{Fate of Stars Formed by This Mechanism}

We discuss the fate of host clouds after the low-mass stars have been formed in the gas shell.  We evaluate the systematic expansion velocity $V$ expected for newborn stars from the expanding velocity of a gas shell at fragmentation in Table~1.   
The escape velocities, $V_{\rm esc}$, of the host clouds are also
 summarized in Table \ref{table:results},
 and are evaluated from the total (baryon $+$ dark-matter) mass
 contained in the SNR shell radius at the fragmentation epoch,
 $M_{\rm T}=M_{\rm sw}(\Omega_0/\Omega_b)$, as
 \begin{equation}
  V_{\rm esc} \simeq \left( \frac{2 G M_{\rm T}}{R}  \right)^{1/2}
   = \left( \frac{2GM_{\rm sw}}{R}\right)^{1/2}
   \left(\frac{\Omega_0}{\Omega_b} \right)^{1/2}
 \end{equation}
 where $M_{\rm sw}$ represents the contained baryon mass.
If $V < V_{\rm esc}$, the second-generation low-mass stars
 remain within the host cloud.    
Since the mass of the host cloud is larger than the minimum mass of fragmentation
 $M_{\rm sw}(\Omega_0/\Omega_b)$, all the models calculated predict that
 newborn stars are bounded.

     Otherwise, the newborn stars will escape from the host cloud and be ejected into intergalactic space.   
This does not occur in our models.
This tendency is consistent with the SPH simulation of SNR formed in low-mass minihalos $M\sim 10^6M_\odot$ 
 by \citet{bromm03}.
That is, in a low energy model with $\varepsilon_0=10^{51}{\rm erg}$ the halo (cloud) is unchanged,
while an SN with $\varepsilon_0=10^{53}{\rm erg}$ disrupts the halo (cloud).

     In our calculations, the fragments are likely to be bounded in the host cloud, because the velocity of the fragments is lower than the escape velocity estimated by the swept mass for  all the models we studied.
     We only investigate the explosion energy of $10^{51} \rm{erg} - 10^{52}$ erg.
If more energetic supernovae, such as pair instability supernovae with
 $\varepsilon_0 > 10^{52} $ erg occur, newborn stars escape from the host
 cloud because their velocities are higher than the escape velocity.
When the SNR escapes from the host cloud without fragmentation, the
     gas, which has large H$_2$ and HD abundances, drifts in the
     intergalactic medium.

\subsection{Metal Abundance} 

\begin{figure}
\plotone{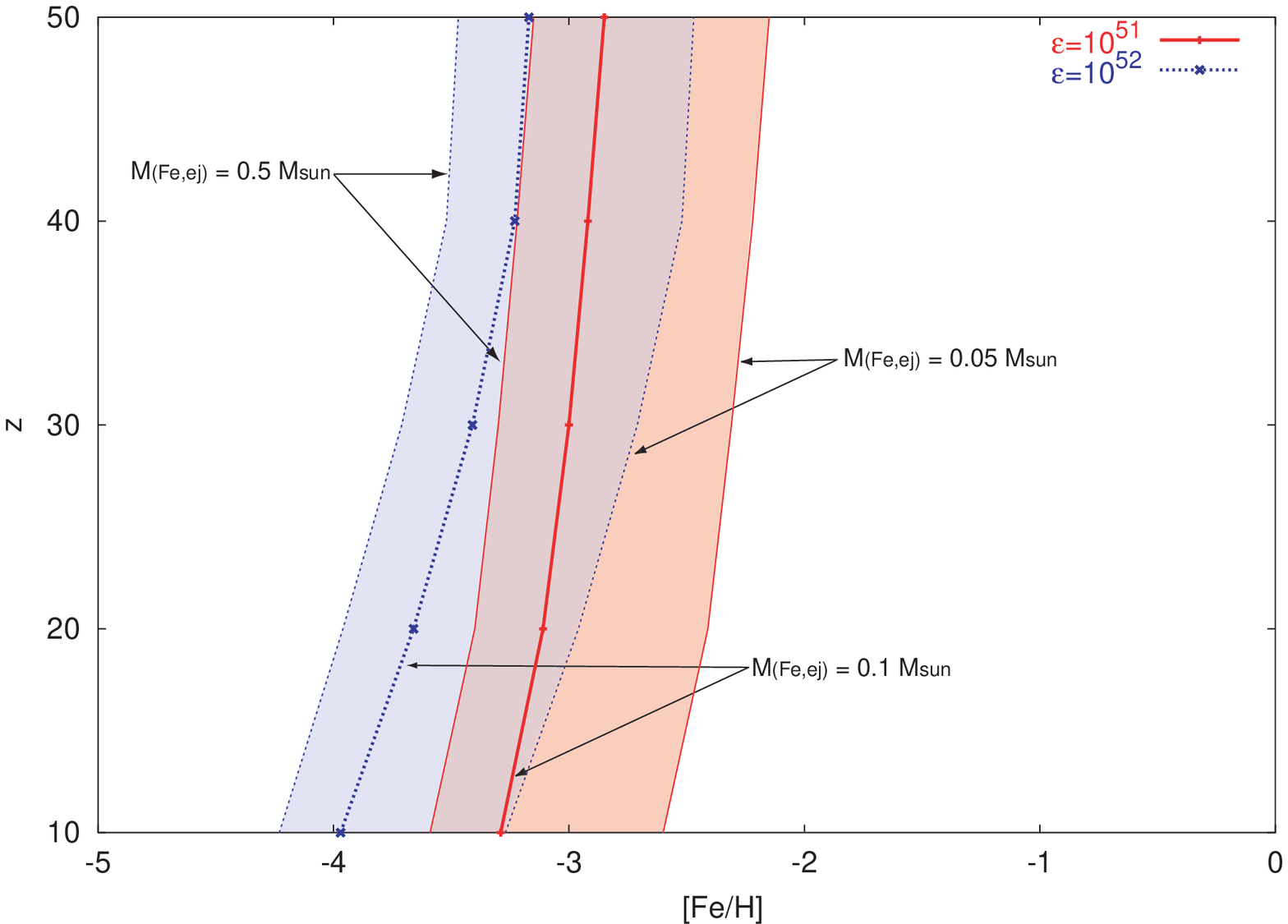}
\caption{
The expected metallicity of [Fe/H] is plotted against the initial $z$.
[Fe/H] is estimated by the ratio of the swept mass and the ejected iron mass.
Ejected iron mass is assumed to be M$_{\rm Fe,ej} = 0.1 \msun $ (thick center lines), 0.05 $\msun$ (left border lines) and 0.5 $\msun$ (right border lines) estimated by \citet{tsujimoto98}.
 Solid and dotted lines represent the model of $\varepsilon_0=10^{51}$ erg and that of $\varepsilon_0=10^{52}$ erg, respectively.}
\label{fig:10}
\end{figure}

In the above, we have assumed the primordial abundances of elements for the gas shell.  
     The fragments, and in particular, the stars formed out of them, are expected to be polluted by metals ejected from the SN, although the degree of metal abundance may vary with the efficiency of mixing between the SN ejecta and the surrounding material.  
     Assuming complete mixing, we can evaluate the metal abundance in the gas shell.
     If the ejected mass of iron, $M_{\rm ej, Fe}$ \citep{tsujimoto98} is mixed with the swept mass, $M_{\rm sw}$, this gives 
\begin{equation}
[{\rm Fe} /{\rm H}] \simeq -2.8 + \log_{10} [(M_{\rm ej, Fe}/0.1 \msun)  
/ (M_{\rm sw} /5 \times 10^4 \msun) ],
\end{equation}
     where the solar abundance of iron is assumed to be equal to $1.16\times10^{-3}$.
     Figure~\ref{fig:10} shows the expected value of [Fe/H].
     From this figure, we found that the next generation stars triggered
     by primordial SNRs have a metallicity of [Fe/H] $\simeq (-2.5) - (-4.5)$. 
     The upper limit corresponds to low explosion energy $\varepsilon_0=10^{51}$ erg and high mass iron ejection $M_{\rm ej,Fe}=0.5 \msun$, while the other limit corresponds to the high explosion energy $\varepsilon_0=10^{52}$ erg and low mass iron ejection $M_{\rm ej,Fe}=0.05 \msun$.
     The actual distributions of metallicity realized in the SNR, however, have to have a scatter around this mean value.  
     It should be pointed out that the metal abundance of the low-mass stars formed by the supernova resembles that of the extremely metal-deficient stars in the Galactic halo [Fe/H]$\sim -3$, discovered by the large-scale survey by \citet{beers92}.
However, the most metal deficient $\feoh = -5.3$ star that has been found, 
 HE0107-5240  \citep{chri02}, has too little metallicity to be explained
 by the mixture of SN ejecta of a first-generation massive star and the
 primordial gas swept in its SNR shell.  
This seems to mean that HE0107-5240 was not formed in a cloud with
 $M_{\rm T}\sim {\rm several} \times 10^6\msun$.

How does the ejecta gas mix with a hundred times more massive cloud composed of primordial gas? 
Consider a cloud with smaller mass as $M_{\rm T} \sim 10^6\msun$.
The SNR shell sweeps all the gas in such a cloud, and the swept shell as well as the gas ejected from the SN
 are ejected into the intracloud space.
In this case, metals mix not with the cloud medium, but with the intracloud medium.
If this cloud belongs to a more massive density perturbation with $M_{\rm T}\sim 10^8\msun$,
 such a system contracts later and forms stars with a hundred times more deficient metallicity
 than the above-mentioned star formed in the cloud $M_{\rm T}\sim {\rm several} \times 10^6\msun$.

     Specifically, our results suggest that the characteristic element abundances of ejecta from a single  population III supernova can be imprinted on the abundance distributions in these extremely metal-poor stars.  
     Their relevance is worthy of further study and if their connection is confirmed, we may gain information on the first collapsed objects in our Universe from the current status of these survivors.

\acknowledgments
We have greatly benefited from discussion with A.~Habe, R.~Nishi, K.~Wada, T.~Matsumoto,
 K.~Omukai, H.~Uehara and T.~Okamoto.  
Numerical calculations were carried out at the Astronomical Data Analysis Center, the National Astronomical Observatory of Japan.
This work is supported in part by the Grants-in-Aid for Science Research
 from MEXT[09640308, 15204010 (MYF), 11640231, 14540233 (KT)].  


\begin{table}
\caption{Initial Conditions and Results}
\begin{center}
\setlength{\tabcolsep}{3pt}
\begin{tabular}{cccc|ccccccccccccc} \hline \hline
\footnotesize{Model} & \footnotesize{$\varepsilon_0$} & \footnotesize{$z$} 
& \footnotesize{$\rho_0$}  & \footnotesize{$t_f$}  & \footnotesize{$r$}  
& \footnotesize{$V $}  & \footnotesize{$V_{\rm esc}$}  
& \footnotesize{$M_{\rm sw}$} & \footnotesize{$T$} &\footnotesize{$T_{\rm hc}$}\
& c$_{\rm s, hc}$
& \footnotesize{[H$_2$/H]} & \footnotesize{[HD/H]} 
\vspace{-2mm}
 \\
& \scriptsize{$(10^{51}{\rm erg})$}  
& &\scriptsize{$({\rm 10^{-25} g\,cm^{-3}})$}&\scriptsize{(Myr)}
& \scriptsize{(pc)} & \scriptsize{$({\rm km\,s^{-1}})$} 
&\scriptsize{$({\rm km\,s^{-1}})$} & \scriptsize{$(10^4M_\odot)$}
& \scriptsize{(K)}&\scriptsize{(K)}& \scriptsize{(km~s$^{-1}$)}&\scriptsize{(log10)} 
&\scriptsize{(log10)} \\
\hline
A10 & 1& 10& 3.51& 45.4& 198& 1.17& 11.1& 17.1& 32& 277& 1.89&-2.72& -4.75\\
B10 & 3& 10& 3.51& 38.7& 251& 1.78& 14.1& 34.9& 36& 303& 1.98&-2.73& -4.83\\
C10 & 5& 10& 3.51& 36.9& 281& 2.12& 15.8& 49.0& 70& 311& 2.01&-2.79& -5.11\\
D10 &10& 10& 3.51& 36.6& 332& 2.61& 18.6& 80.4& 98& 312& 2.01&-2.83& -5.41\\
\hline
A20 & 1& 20& 24.4& 14.7&  90& 1.73& 13.2& 11.1& 53& 223& 1.70&-2.80& -5.08\\
B20 & 3& 20& 24.4& 12.2& 111& 2.43& 16.4& 20.8& 69& 248& 1.83&-2.80& -5.20\\
C20 & 5& 20& 24.4& 10.9& 122& 2.99& 17.9& 27.4& 83& 262& 1.85&-2.80& -5.29\\
D20 &10& 20& 24.4& 10.9& 137& 3.94& 20.2& 39.2&109& 282& 1.91&-2.81& -5.46\\
\hline

A30 & 1& 30& 78.5& 8.42&  56& 1.75& 14.9& 8.75& 75& 199& 1.61&-2.78& -5.01\\
B30 & 3& 30& 78.5& 6.55&  68& 2.49& 18.1& 15.6& 79& 228& 1.72&-2.78& -5.09\\
C30 & 5& 30& 78.5& 4.73&  70& 4.04& 18.6& 16.9& 87& 278& 1.91&-2.78& -5.26\\
D30 &10& 30& 78.5& 3.79&  77& 5.62& 20.3& 22.3&120& 321& 2.04&-2.78& -5.51\\
\hline

A40 & 1& 40& 182& 5.85&  40& 1.85& 16.0& 7.16&  99& 194& 1.54& -2.76& -5.17\\
B40 & 3& 40& 182& 5.10&  49& 2.51& 20.3& 14.3& 102& 207& 1.64&-2.76& -5.19\\
C40 & 5& 40& 182& 3.17&  50& 4.04& 20.3& 14.4& 108& 276& 1.89&-2.76& -5.28\\
D40 &10& 40& 182& 1.57&  51& 9.02& 20.2& 14.5& 151& 656& 2.92&-2.77& -5.73\\

\hline
A50 & 1& 50& 349& 4.40&  30& 1.97& 17.0& 6.07& 124& 193& 1.58&-2.73& -5.40\\
B50 & 3& 50& 349& 3.82&  38& 2.52& 21.4& 12.2& 127& 206& 1.64&-2.74& -5.42\\
C50 & 5& 50& 349& 2.73&  39& 4.18& 21.6& 12.7& 132& 260& 1.84&-2.74& -5.46\\
D50 &10& 50& 349& 1.08&  39& 9.09& 21.6& 12.7& 154& 567& 2.71&-2.74& -5.66\\
\hline
\end{tabular}
\label{table:results}
\end{center}
\tablenotetext{}{
Symbols $t_f$, $r$, $V$, $V_{\rm esc}$, $M_{\rm sw}$, $T$, $T_{\rm hc}$, $c_{\rm s, hc}$,
[H$_2$/HD] and [HD/H] represent the elapsed time, the radius of the shell, 
the expansion speed, the escape velocity, the swept mass, temperature of the shell, temperature of the ambient medium, sound speed of the ambient medium, and H$_2$ and  HD fractional abundances in the logarithmic scale, respectively.
}
\end{table}

\clearpage
\appendix
\section{Radiative Cooling Process}
As the radiative cooling process,  we include
 the inverse Compton process owing to the cosmic background radiation, and line cooling of H, He, H$_2$, and HD.
Here in this Appendix, we briefly summarize the cooling rate.
    The unit of the cooling rate is erg cm$^{-3}$ s$^{-1}$ and that of the temperature is K.
\begin{enumerate}
\item Inverse Compton cooling \citep{ikeuchi86} :
   \begin{equation}
      \Lambda_{\rm{ic}} = 5.41 \times 10^{-32} \ (1+z)^4 \  
      \left( \frac{T}{10^4} \right) \ n_{\rm{e}}.
   \end{equation}
\item Helium cooling 
   \begin{enumerate}
   \item Collisional ionization cooling \citep{cen92} :
      \begin{eqnarray}
         \Lambda_{\rm{He,cl}} &=& 9.38 \times 10^{-22} \ T^{1/2} \ 
         \left[ 1+ \left( \frac{T}{10^5} \right)^{1/2} \right]^{-1} \
          {\rm{exp}} (-285335.4/T) 
         \ n_{\rm{e}} \ n_{\rm{He}}, \\
         \Lambda_{\rm{He^+,cl}} &=& 4.95 \times 10^{-22} \ T^{1/2} \
         \left[ 1+ \left( \frac{T}{10^5} \right)^{1/2} \right]^{-1} 
         {\rm{exp}} (-631515/T) \ n_{\rm{e}} \ n_{\rm{He^+}}, \\
         \Lambda_{\rm{He^{+},cl}}^{\prime} &=& 5.01 \times 10^{-27} \ T^{-0.1687} \
         \left[ 1+ \left( \frac{T}{10^5} \right)^{1/2} \right]^{-1} 
         {\rm{exp}} (-55338/T) \ n_{\rm{e}}^2 \ n_{\rm{He^+}}.
      \end{eqnarray}
   \item Recombination cooling \citep{cen92} :
      \begin{eqnarray}
         \Lambda_{\rm{He^+,re}} &=&1.55\times 10^{-26} \ T^{0.3647} 
         \ n_{\rm{e}} \ n_{\rm{He^+}}, \\
         \Lambda_{\rm{He^{++},re}} &=& 3.48 \times 10^{-26} T^{1/2} 
         \left( \frac{T}{10^3}\right)^{-0.2} /
         \left[ 1 + \left( \frac{T}{10^6}\right)^{0.7} \right] 
         n_{\rm{e}} \ n_{\rm{He^{++}}}.
      \end{eqnarray}   
   \item Collisional excitation cooling \citep{cen92} :
      \begin{eqnarray}
         \Lambda_{\rm{He^+,ex}} &=& 5.54 \times 10^{-17} \ T^{-0.397}
         \left[  1+ \left( \frac{T}{10^5} \right)^{1/2} \right]^{-1} \
         {\rm{exp}} (-473638/T) \ n_e \ n_{\rm He^{+}}, \\ 
         \Lambda_{\rm{He^{++},ex}} &=& 9.10 \times 10^{-27} \ T^{-0.1687}
         \left[  1+ \left( \frac{T}{10^5} \right)^{1/2} \right]^{-1} \
         {\rm{exp}} (-13179/T) \ n_e^2 \ n_{\rm He^{+}}.  
      \end{eqnarray}  
   \end{enumerate}
\item Hydrogen cooling
   \begin{enumerate}
   \item Collisional ionization cooling \citep{cen92} :
      \begin{eqnarray}
         \Lambda_{\rm{H,cl}} =1.27\times 10^{-21} \ T^{1/2}
         \left[1+ \left( \frac{T}{10^5} \right)^{1/2}\right]^{-1} \ 
         {\rm{exp}}(-157809.1/T) \ n_{\rm{e}} \ n_{\rm{H}}. 
      \end{eqnarray}
   \item Recombination cooling \citep{cen92} :
      \begin{eqnarray}
         \Lambda_{\rm{H,re}}=8.70 \times 10^{-27} \ T^{1/2} 
         \left(\frac{T}{10^3} \right)^{-0.2}
         /\left[1+\left( \frac{T}{10^6}\right)^{0.7} \right]
         n_{\rm{e}} \ n_{\rm{H^+}}.
      \end{eqnarray}
   \item Collisional excitation cooling \citep{cen92}:
      \begin{eqnarray}
         \Lambda_{\rm{H,ex}} = 7.5\times10^{-19}
         \left[1+ \left( \frac{T}{10^5} \right)^{1/2} 
         \right]^{-1} {\rm{exp}}(-118348/T) \ n_{\rm{e}} \ n_{\rm H}.
      \end{eqnarray}
   \end{enumerate} 
\item Molecule Hydrogen cooling:
$\Lambda_{\rm H_2}$ is taken from the table of Flower \etal (2000).  

\item HD cooling:
$\Lambda_{\rm HD}$ is taken from the table of Flower \etal (2000).  

\item Effect of the CMB radiation:  
\begin{equation}
\Lambda_{\rm CMB} =  \Lambda_{\rm H_2}(T_{\rm CMB}) + \Lambda_{\rm HD} (T_{\rm CMB} ), 
\end{equation}
where $T_{\rm CMB}$ is taken as 2.73(1+z).
\end{enumerate}

\newpage
\section{Chemical Reactions}
We include the chemical reactions of 12 species: 
H, H$^+$, H$^-$, He, He$^+$, He$^{++}$, H$_2$, D,  D$^+$, HD, HD$^+$, and e$^-$.
In this Appendix we summarize the reactions we adopt.
\begin{eqnarray}
\frac{dn_{\rm{H}}}{dt}
 &=& k_2\, n_{\rm{H^+}}\, n_{\rm{e}} 
  +  2\, k_{12} \, n_{\rm{H_2}} \, n_{\rm{e}}
  +  k_{13}\, n_{\rm{H^-}} \, n_{\rm{e}} 
  +  2\, k_{14}\, n_{\rm{H^-}} \,  n_{\rm{H^+}} 
  +  k_{20}\, n_{\rm{D}} \,  n_{\rm{H^+}} \nonumber \\
 &+&  k_{22}\, n_{\rm{D}} \,  n_{\rm{H_2}} 
  + k_9\, n_{\rm{H}} \,  n_{\rm{H}} \,  n_{\rm{H}} 
  +  3\,k_{10}\, n_{\rm{H2}}  \, n_H 
  +  k_{11}\, n_{\rm{H_2}}  \, n_{\rm{H^+}} 
  +  2\,k_{16}\, n_{\rm{H_2}}  \, n_{\rm{H_2}} \nonumber \\
 &+&  k_{17}\, n_{\rm{H}}  \, n_{\rm{H}}  
  - k_1\, n_{\rm{H}} \,  n_{\rm{e}}-k_7 \,  n_{\rm{H}} \,  n_{\rm{e}}
  -  k_8\, n_{\rm{H}}  \, n_{\rm{H^-}} 
  -  k_{21}\, n_{\rm{D^+}} \,  n_{\rm{H}} \nonumber \\
 &-& k_{24}\, n_{\rm{HD^+}} \, n_{\rm{H}} 
  -  k_{28}\, n{_{\rm{D^+}}}  \, n_{\rm{H}} 
  - 3 k_9\, n_{\rm{H}}  \, n_{\rm{H}}  \, n_{\rm{H}} 
  -  k_{10}\, n_{\rm{H_2}}  \, n_{\rm{H}} \nonumber \\
 &-&  2 \, k_{15}\, n_{\rm{H}}  \, n_{\rm{H}}  \, n_{\rm{H_2}}
  -  2 \, k_{17}\, n_{\rm{H}}  \, n_{\rm{H}}, \\
\frac{dn_{\rm{H^+}}}{dt} 
 &=&  k_1\,  n_{\rm{H}}\,  n_{\rm{e}} 
  + k_{21}\,  n_{\rm{D^+}}\,  n_{\rm{H}}
  + k_{23}\,  n_{\rm{HD^+}}\,  n_{\rm{H^+}}
  +  k_{24}\,  n_{\rm{D^+}}\, n_{\rm{H_2}}
  +  k_{17}\,  n_{\rm{H}}\,  n_{\rm{H}} \nonumber \\
 &-&  k_2\,  n_{\rm{H^+}}\, n_{\rm{e}} 
  - k_{11}\, n_{\rm{H_2}}\, n_{\rm{H^+}}
  - k_{14}\, n_{\rm{H^-}}\, n_{\rm{H^+}}
  - k_{20}\, n_D\,  n_{\rm{H^+}}  
  -  k_{26}\, n_{\rm{HD}}\, n_{\rm{H^+}} \nonumber \\
 &-& k_{27}\, n_{\rm{D}}\, n_{\rm{H^+}}, \\
\frac{dn_{\rm{H^-}}}{dt}
  &=& k_7\, n_{\rm{H}}\, n_{\rm{e}} 
   - k_8\, n_{\rm{H}}\, n_{\rm{H^-}}
   - k_{13}\, n_{\rm{H^-}}\, n_{\rm{e}} 
   - k_{14}\, n_{\rm{H^-}}\, n_{\rm{H^+}}, \\
\frac{dn_{\rm{He}}}{dt}
   &=&  k_4\,  n_{\rm{He^+}}\,  n_{\rm{e}} 
    - k_3\,  n_{\rm{He}}\,  n_{\rm{e}}, \\
\frac{dn_{\rm{He^+}}}{dt}
   &=& k_3\,  n_{\rm{He}}\,  n_{\rm{e}} 
    + k_6\,  n_{\rm{He^{++}}}\,  n_{\rm{e}} 
    - k_4\,  n_{\rm{He^+}}\,  n_{\rm{e}}
    - k_5\,  n_{\rm{He^{+}}}\, n_{\rm e}, \\
\frac{dn_{\rm{He^{++}}}}{dt}
   &=&\, k_5\,  n_{\rm{He^+}}\,  n_e 
    -  k_6\,  n_{\rm{He^{++}}}\,  n_{\rm{e}}, \\
\frac{dn_{\rm{H_2}}}{dt}
    &=&\,k_8 n_{\rm{H}}\, n_{\rm{H^-}} 
     +   k_{26}\,n_{\rm{HD}}\,n_{\rm{H^+}} 
     +  k_9\, n_{\rm{H}}\, n_{\rm{H}}\, n_{\rm{H}} 
     +   2\, k_{15}\, n_{\rm{H}}\, n_{\rm{H}}\, n_{\rm{H_2}}\nonumber \\ 
    &+&   k_{16}\,n_{\rm{H_2}}\,n_{\rm{H_2}} 
     -  k_{11}\,n_{\rm{H_2}}\,n_{\rm{H^+}} 
     -   k_{12}\,n_{\rm{H_2}}\,n_e
     -   k_{22}\,n_{\rm{D}}\,n_{\rm{H_2}}
     -   k_{24}\,n_{\rm{D^+}}\,n_{\rm{H_2}} \nonumber \\
    &-&  k_{10}\, n_{\rm{H_2}}\,n_{\rm{H}} 
     -   k_{15}\, n_{\rm{H}}\, n_{\rm{H}}\, n_{\rm{H_2}} 
     -   2\,k_{16}\, n_{\rm{H_2}}\, n_{\rm{H_2}}, \\
\frac{dn_{\rm{D}}}{dt}
    &=& k_{18}\, n_{\rm{D^+}}\, n_{\rm{e}} 
     +k_{21}\,  n_{\rm{D^+}}\, n_{\rm{H}} 
     + k_{25}\,  n_{\rm{HD}}\, n_{\rm{H}}  
    - k_{19}\, n_{\rm{D}}\, n_{\rm{e}} \nonumber \\
    &-&k_{20}\,  n_{\rm{D}}\, n_{\rm{H^+}}
     -k_{22}\, n_{\rm{D}}\, n_{\rm{H_2}} \\
\frac{dn_{\rm{D^+}}}{dt}
   &=& k_{20}\, n_{\rm{D}}\, n_{\rm{H^+}}
    +  k_{26}\, n_{\rm{HD}}\, n_{\rm{H^+}}
    -  k_{18}\,  n_{\rm{D^+}}\, n_{\rm{e}}
    -  k_{21}\,  n_{\rm{D^+}}\, n_{\rm{H}} \nonumber \\
   &-& k_{24}\,  n_{\rm{D^+}}\, n_{\rm{H_2}}
    -  k_{28}\, n_{\rm{D^+}}\, n_{\rm{H}}, \\
\frac{dn_{\rm{HD}}}{dt}
    &=& k_{22}\,  n_{\rm{D}}\,  n_{\rm{H_2}}
     +  k_{23}\,  n_{\rm{HD^+}}\,  n_{\rm{H}} 
     +  k_{24}\, n_{\rm{D^+}}\, n_{\rm{H_2}}
     -  k_{26}\, n_{\rm{HD}}\,  n_{\rm{H^+}}, \\ 
\frac{dn_{\rm{HD^+}}}{dt}
    &=& k_{27}\, n_{D}\, n_{\rm{H^+}}
     +  k_{28}\, n_{\rm{D^+}}\, n_{\rm{H}} 
     -  k_{23}\, n_{\rm{HD^+}}\, n_{\rm{H}}. 
\end{eqnarray}

The reaction rates for the above reactions are taken 
 from \citet{pal83}, \citet{shapiro87}, \citet{abel97},  \citet{gal98} and
 \citet{sta98}.

\begin{table}
\setlength{\tabcolsep}{5pt}
\renewcommand{\arraystretch}{0.4}
\begin{center}
\begin{tabular}{lllc}
\hline \hline
 & reaction & rate \ (cm$^{-3} $s$^{-1}  $)   & reference \\
\hline \\
(1)&  H \ +\ e $\rightarrow$ H$^+$ \ +\ 2 e & $k_1=\ $exp$[\ - \ 32.71396786$ & 1 \\
  && \hspace{17mm} $ +\ 13.536556 \times $ln$(T/$ev$)     $\\
  && \hspace{17mm} $ -\ 5.73932875 \times $ln$(T/$ev$)^2 $\\
  && \hspace{17mm} $ +\ 1.56315498 \times $ln$(T/$ev$)^3  $\\ 
  && \hspace{17mm} $ -\ 0.2877056\times $ln$(T/$ev$)^4    $\\
  && \hspace{17mm} $ +\ 3.48255977\times10^{-2} \times $ln$(T/$ev$)^5    $\\ 
  && \hspace{17mm} $ -\ 2.63197617\times10^{-3} \times $ln$(T/$ev$)^6    $\\
  && \hspace{17mm} $ +\ 1.11954395\times10^{-4} \times $ln$(T/$ev$)^7    $\\ 
  && \hspace{17mm} $ -\ 2.03914985\times10^{-6} \times $ln$(T/$ev$)^8 \ ] $\\
\\
(2)&  H$^+$  + e $\rightarrow$ H + $\gamma$  &  $k_2=\ $exp$[\ -\ 28.6130338 $ & 1\\
  && \hspace{17mm} $ -\ 0.72411256 \times $ln$(T/$ev$)$ \\
  && \hspace{17mm} $ -\ 2.02604473 \times 10^{-2} \times $ln$(T/$ev$)^2$  \\
  && \hspace{17mm} $ -\ 2.38086188 \times 10^{-3} \times $ln$(T/$ev$)^3 $ \\ 
  && \hspace{17mm} $ -\ 3.21260521 \times 10^{-4} \times $ln$(T/$ev$)^4 $ \\
  && \hspace{17mm} $ -\ 1.42150291 \times 10^{-5} \times $ln$(T/$ev$)^5 $ \\ 
  && \hspace{17mm} $ +\ 4.98910892 \times 10^{-6} \times $ln$(T/$ev$)^6 $ \\
  && \hspace{17mm} $ +\ 5.75561414 \times 10^{-7} \times $ln$(T/$ev$)^7 $ \\ 
  && \hspace{17mm} $ -\ 1.85676704 \times 10^{-8} \times $ln$(T/$ev$)^8 $ \\ 
  && \hspace{17mm} $ -\ 3.07113524 \times 10^{-9} \times $ln$(T/$ev$)^9 $ \ ]\\ 
\\
(3)&  He + e $\rightarrow$ He$^+$ + 2 e & $k_3 =\ $exp$[ \ -44.09864886$ & 1 \\ 
  && \hspace{17mm} $ +\ 23.91596563 \times $ln$(T/$ev$)  $\\  
  && \hspace{17mm} $ -\ 10.7532302 \times $ln$(T/$ev$)^2 $\\
  && \hspace{17mm} $ +\ 3.05803875 \times $ln$(T/$ev$)^3 $\\ 

  && \hspace{17mm} $ -\ 0.56851189\times $ln$(T/$ev$)^4 $\\
  && \hspace{17mm} $ +\ 6.79539123\times10^{-2} \times $ln$(T/$ev$)^5 $\\ 
  && \hspace{17mm} $ -\ 5.00905610\times10^{-3} \times $ln$(T/$ev$)^6 $\\
  && \hspace{17mm} $ +\ 2.06723616\times10^{-4} \times $ln$(T/$ev$)^7 $\\ 
  && \hspace{17mm} $ -\ 3.694916141\times10^{-6} \times $ln$(T/$ev$)^8\ ] $\\
\\
(4)&He$^+$ + e $\rightarrow$ He + $\gamma$  &$k_4 = 3.925\times 10^{-13} 
\times (T/$ev$)^{-0.6353}$ & 1\\
\\
(5)&  He$^+$ + e $\rightarrow$ He$^{++}$ + 2e & $k_5 =\ $exp$[ \ -68.71040990 $& 1\\
  && \hspace{17mm} $ +\ 43.93347633 \times $ln$(T/$ev$)  $\\ 
  && \hspace{17mm} $ -\ 18.4806699 \times $ln$(T/$ev$)^2 $\\
  && \hspace{17mm} $ +\ 4.70162649 \times $ln$(T/$ev$)^3 $\\ 
  && \hspace{17mm} $ -\ 0.76924663 \times $ln$(T/$ev$)^4 $\\
  && \hspace{17mm} $ +\ 8.113042 \times10^{-2} \times $ln$(T/$ev$)^5 $\\ 
  && \hspace{17mm} $ -\ 5.32402063 \times10^{-3} \times $ln$(T/$ev$)^6 $\\
  && \hspace{17mm} $ +\ 1.97570531 \times10^{-4} \times $ln$(T/$ev$)^7 $\\ 
  && \hspace{17mm} $ -\ 3.16558106 \times10^{-6} \times $ln$(T/$ev$)^8\ ] $\\
\\
(6)&  He$^{++}$ + e $\rightarrow$ He$^+$ + $\gamma$& $
k_6 = 3.36  10^{-10}\  T^{-1/2} 
\left( \frac{T}{1000}\right)^{-0.2} \times 
\left[1+\left(\frac{T}{10^6}\right)^{0.7} \right]^{-1} $
  & 1\\
\\
(7)&  H  + e $\rightarrow$ H$^-$ + $\gamma$ &  
$k_7 = \left\{ \begin{array}{l} 1.0 \times 10^{-18} \ 
 T \hspace{15mm} T<1.5\times 10^4 \ $K$ \\ \\
   $dex$ [\ -\ 14.10  \hspace{16mm} T>1.5\times 10^4 \ $K$ \\
 \hspace{9mm} +\ 0.1175\  $log T$ \\
 \hspace{9mm} -\ 9.813 \times10^{-3} \ ($log$T)^2 \ ]  
 \end{array} \right. $
  & 2\\
\\
\end{tabular}
\end{center}
\end{table}

\begin{table}
\setlength{\tabcolsep}{5pt}
\renewcommand{\arraystretch}{0.4}
\begin{center}
\setlength{\tabcolsep}{2pt}
\begin{tabular}{lllc}
\hline \hline
 & reaction & rate \ (cm$^{-3} $s$^{-1}  $)  & reference \\
\hline \\
(8)&  H + H$^-$ $\rightarrow$ H$_2$ + e &
 $ k_8 = 
\left\{ \begin{array}{l}
 $exp$[\ -\ 20.06913897  \hspace{10mm}  T>1160\ $K$ \\
   \hspace{9mm} +\ 0.22898 \times $ln$(T/$ev$) \\   
   \hspace{9mm} +\ 3.5998377 \times10^{-2} \times $ln$(T/$ev$)^2   \\
   \hspace{9mm} -\ 4.55512 \times10^{-3} \times $ln$(T/$ev$)^3    \\ 
   \hspace{9mm} -\ 3.10511544 \times10^{-4} \times $ln$(T/$ev$)^4 \\
   \hspace{9mm} +\ 1.0732940 \times10^{-4} \times $ln$(T/$ev$)^5  \\ 
   \hspace{9mm} -\ 8.36671960 \times10^{-6} \times $ln$(T/$ev$)^6 \\
   \hspace{9mm} +\ 2.23830623 \times10^{-7} \times $ln$(T/$ev$)^7
    \\ \\
1.428\times 10^{-9} \hspace{20mm} T<1160\ $K$ 
\end{array} \right. $
   & 1\\ 
\\
(9)&  H + H + H $\rightarrow$ H$_2$ + H       & 
$ k_9 = 5.5\times10^{-29} \times T^{-1} $ & 3\\
\\
(10)& H$_2$ + H $\rightarrow$ H + H + H       &  
$k_{10} = 6.5 \times 10^{-7} T^{-1/2} \ $exp$(-52000/T)
 $ & 3\\
  && \hspace{5mm} $\times [1-$exp$(-6000/T)] $ \\
\\
(11)& H$_2$ +H$^+$ $\rightarrow$ H$_2^
+$ + H& $k_{11} = \ $exp$ [-\ 24.24914687$ & 1 \\
  && \hspace{17mm} $ +\ 3.4008244 \times\ln (T/$ev$)$ & 1\\
  && \hspace{17mm} $ -\ 3.89800396 \times\ln (T/$ev$)^2 $\\
  && \hspace{17mm} $ +\ 2.04558782 \times\ln (T/$ev$)^3 $\\ 
  && \hspace{17mm} $ -\ 0.541618285 \times\ln (T/$ev$)^4 $\\
  && \hspace{17mm} $ +\ 8.41077503 \times10^{-2} \times\ln (T/$ev$)^5 $\\ 
  && \hspace{17mm} $ -\ 7.87902615 \times10^{-3} \times\ln (T/$ev$)^6 $\\
  && \hspace{17mm} $ +\ 4.13839842 \times10^{-4} \times\ln (T/$ev$)^7 $\\ 
  && \hspace{17mm} $ -\ 9.3634588 \times10^{-6} \times\ln (T/$ev$)^8\ ] $\\
\\
(12)& H$_2$ +\ e $\rightarrow$ 2H + e           &  
$k_{12} = 5.6\times10^{-11}  T^{1/2} \ $exp$(-102124/T) $ &  1\\
\\
(13)& H$^-$ + e $\rightarrow$ H + 2e&$k_{13} = $exp$[-\ 18.01849334$ & 1\\
  && \hspace{17mm} $ +\ 2.3608522  \times\ln (T/$ev$)$ \\
  && \hspace{17mm} $ -\ 0.28274430   \times\ln (T/$ev$)^2 $\\
  && \hspace{17mm} $ +\ 1.62331664 \times10^{-2} \times \ln (T/$ev$)^3 $\\ 
  && \hspace{17mm} $ -\ 3.36501203 \times10^{-2} \times \ln (T/$ev$)^4 $\\
  && \hspace{17mm} $ +\ 1.17832978 \times10^{-2} \times \ln (T/$ev$)^5 $\\ 
  && \hspace{17mm} $ -\ 1.65619470 \times10^{-3} \times \ln (T/$ev$)^6 $\\
  && \hspace{17mm} $ +\ 1.06827520 \times10^{-4} \times \ln (T/$ev$)^7 $\\ 
  && \hspace{17mm} $ -\ 2.63128581 \times10^{-6} \times \ln (T/$ev$)^8\ ] $\\
\\
(14)& H$^-$ + H$^+$ $\rightarrow$ 2H           &  
$k_{14} = 7\times 10^{-8}  \left( \frac{T}{100} \right)^{-1/2} $ &  1\\
\\
(15)& H + H + H$_2$ $\rightarrow$ H$_2$ + H$_2$   &  
$k_{15}=6.875\times 10^{-30} T^{-1} $ & 3\\
\\
(16)& H$_2$ + H$_2$ $\rightarrow$ H + H + H$_2$   &  
$k_{16} = 8.0\times 10^{-8} T^{-1/2} \ $exp$(-52000/T) $ &3\\
&& \hspace{15mm} $\times \left[ 1-{\rm exp}(-6000/T) \right] $ \\
\\
(17)& H + H  $\rightarrow$ H$^+$ + e + H      & 
$k_{17} =9.86 \times 10^{-15} T^{0.5} $exp$(-158000/T) $ &3\\
\\
(18)& D$^+$ + e  $\rightarrow$ D + $\gamma$     &  
$k_{18}=  3.6 \times 10^{-12}(T/300) $  &4\\
\\
(19)& D + e $\rightarrow$ D$^-$ +  $\gamma$       & 
$k_{19}=3.0 \times 10^{-16}(T/300)^{0.95} $exp$(-T/9320) $ &4\\
\\
(20)& D + H$^+$ $\rightarrow$ D$^+$ + H         & 
 $k_{20} = 3.7\times 10^{-10}  T^{0.28}  $exp$(-43/T) $   &4\\
\end{tabular}
\end{center}
\end{table}

\begin{table}
\setlength{\tabcolsep}{5pt}
\renewcommand{\arraystretch}{0.4}
\begin{center}
\setlength{\tabcolsep}{2pt}
\begin{tabular}{lllc}
\hline \hline
 & reaction & rate \ (cm$^{-3} $s$^{-1}  $)  & reference \\
\hline \\
(21)& D$^+$ + H $\rightarrow$ D + H$^+$         & 
$k_{21} = 3.7 \times 10^{-10} \ T^{0.28} $  &4\\
\\
(22)& D + H$_2$ $\rightarrow$ H + HD          & 
 $k_{22}= 9\times10^{-11} \ $exp$(-3876/T) $   &4\\
\\
(23)& HD$^+$ + H $\rightarrow$ H$^+$ + HD        &  
$k_{23}= 6.4 \times 10^{-10}$  &4\\
\\
(24)& D$^+$ + H$_2$ $\rightarrow$ H$^+$ + HD      &
 $k_{24}=2.1\times10^{-9}$ &4\\
\\
(25)& HD + H  $\rightarrow$ H$_2$ + D         & 
 $k_{25} = 3.2 \times 10^{-11} \ $exp$(-3624/T) $  &4\\
\\
(26)& HD + H$^+$ $\rightarrow$ H$_2$ + D$^+$      &
$ k_{26} = 1.0 \times 10^{-9} \ $exp$(-464/T)  $  &4\\
\\
(27)& D + H$^+$ $\rightarrow$ HD$^+$ + $\gamma$   & 
$ k_{27}=$ dex $[-\ 19.38-1.523 \times\log T   $ &5\\
  && $\hspace{20mm} +\ 1.118  \times(\log T)^2$ \\
  && $\hspace{20mm} -\ 0.1269 \times(\log T)^3\ ]$ \\
\\
(28)& D$^+$ + H $\rightarrow$ HD$^+$ + $\gamma$   &  
$k_{28}= $dex$[-\ 19.38-1.523 \times\log T $ & 5\\
  && $\hspace{17mm} + \ 1.118\times(\log T)^2 - 0.1269 \times(\log T)^3]$ \\
\hline
\end{tabular}

\end{center}
\caption{
References.---1. \citet{abel97}; 2. \citet{shapiro87}; 3. \citet{pal83};
 4. \citet{gal98}; 5. \citet{sta98}.
}
\end{table}

\clearpage

\end{document}